\documentclass[journal, onecolumn, 12pt]{IEEEtran}
\renewcommand{\baselinestretch}{2.0}
\usepackage{cite}
\usepackage{ifpdf}
\usepackage{array}
\ifpdf
\usepackage{graphicx}
\usepackage[svgnames]{xcolor}
\else
\usepackage[dvipdfmx]{graphicx}
\usepackage[dvipdfmx,svgnames]{xcolor}
\fi
\usepackage{amsmath,amssymb,amscd}
\usepackage{verbatim}
\usepackage{algorithm}
\usepackage{algpseudocode}

\begin{document}
\renewcommand{\baselinestretch}{1.55}
\title{Layered Downlink Precoding for C-RAN Systems with Full Dimensional MIMO}

\author{\large Jinkyu Kang, Osvaldo Simeone, Joonhyuk Kang and Shlomo Shamai (Shitz)
\thanks{Jinkyu Kang and Joonhyuk Kang are with the Department of Electrical Engineering, Korea Advanced Institute of Science and Technology (KAIST) Daejeon, South Korea (Email: kangjk@kaist.ac.kr and jhkang@ee.kaist.ac.kr).

O. Simeone is with the Center for Wireless Communications and Signal Processing Research (CWCSPR), ECE Department, New Jersey Institute of Technology (NJIT), Newark, NJ 07102, USA (Email: osvaldo.simeone@njit.edu). 

S. Shamai (Shitz) is with the Department of Electrical Engineering, Technion, Haifa, 32000, Israel (Email: sshlomo@ee.technion.ac.il).
}
}
\maketitle
\renewcommand{\baselinestretch}{2.0}
\begin{abstract}
The implementation of a Cloud Radio Access Network (C-RAN) with Full Dimensional (FD)-MIMO is faced with the challenge of controlling the fronthaul overhead for the transmission of baseband signals as the number of horizontal and vertical antennas grows larger. This work proposes to leverage the special low-rank structure of FD-MIMO channel, which is characterized by a time-invariant elevation component and a time-varying azimuth component, by means of a layered precoding approach, so as to reduce the fronthaul overhead. According to this scheme, separate precoding matrices are applied for the azimuth and elevation channel components, with different rates of adaptation to the channel variations and correspondingly different impacts on the fronthaul capacity. Moreover, we consider two different Central Unit (CU) - Radio Unit (RU) functional splits at the physical layer, namely the conventional C-RAN implementation and an alternative one in which coding and precoding are performed at the RUs. Via numerical results, it is shown that the layered schemes significantly outperform conventional non-layered schemes, especially in the regime of low fronthaul capacity and large number of vertical antennas.
\end{abstract}

\section{Introduction}
The cloud radio access network (C-RAN) architecture consists of multiple radio units (RUs) connected via fronthaul links to a central unit (CU) that implements the protocol stack of the RUs, including baseband processing \cite{ChinaMobile, Checko}. C-RAN enables a significant reduction in capital and operating expenses, as well as an enhanced spectral efficiency by means of joint interference management at the physical layer across all connected RUs. Nevertheless, it is well recognized that the performance of this architecture is limited by the capacity and latency constraints of the fronthaul network connecting RUs and CU \cite{ChinaMobile, Checko, Samardzija12TWC, Park14SPMAG}.

In a standard C-RAN implementation, the fronthaul links carry digitized baseband signals. Hence, the bit rate required for a fronthaul link is determined by the quantization and compression operations applied to the baseband signals prior to transmission on the fronthaul links. As such, the fronthaul rate is proportional to the signal bandwidth, to the oversampling factor, to the resolution of the quantizer/compressor, and to the number of antennas \cite{Dotsch13Bell}. The fronthaul bit rate can be reduced by implementing alternative functional splits between CU and RU, whereby some baseband functionalities are implemented at the RU \cite{Wubben14SPMAG, Rost2015WC, DeLaOliva2015WC}.

As a concurrent trend in the evolution of wireless networks, in the 3rd generation partnership project (3GPP) long term evolution (LTE) Release-13, three-dimensional (3D)-MIMO, where base stations are equipped with two-dimensional rectangular antenna arrays, has been intensely discussed as a promising tool to boost spectral efficiency \cite{Kuo15WC, Nam13COMMAG}. 3D-MIMO technology is classified into three categories, namely, vertical sectorization (VS), elevation beamforming (EB), and Full-Dimensional MIMO (FD-MIMO) in order of complexity. The VS scheme splits a sector of cellular coverage into multiple sectors by means of different electrical downtilt angles. With the EB approach, instead, users are supported by predetermined or adaptive beams in the elevation direction. Finally, in FD-MIMO, the spatial diversity provided by vertical and horizontal antennas is leveraged jointly to serve multiple users using multiuser-MIMO techniques.

\begin{figure}[t]
\centering
\vspace{-0.5cm}
\includegraphics[width=14cm]{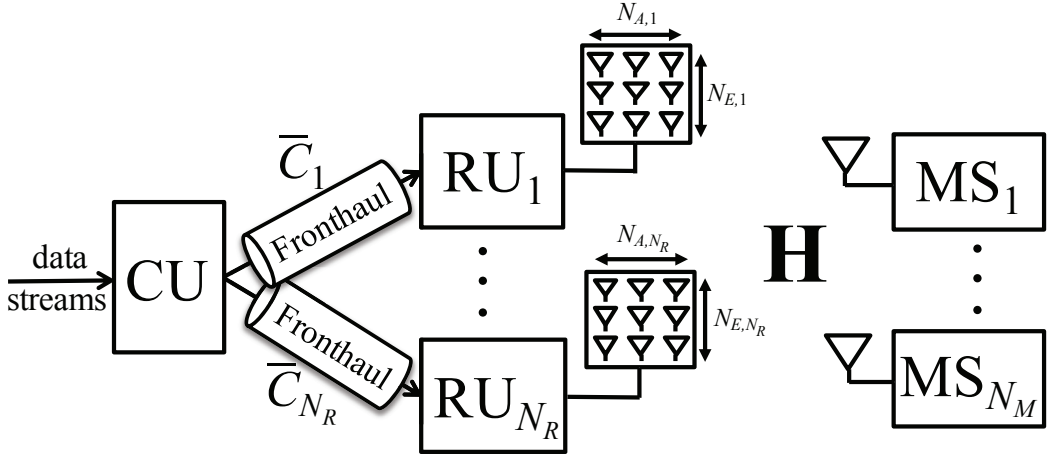}
\vspace{-0.5cm}
\caption{Downlink of a C-RAN system with FD-MIMO.} 
\vspace{-0.5cm}
\label{fig:fig1}
\end{figure}

Endowing RUs with two-dimensional arrays in a C-RAN system (see Fig. \ref{fig:fig1}), while promising from a spectral efficiency perspective, creates significant challenges in terms of fronthaul overhead as the number of antennas grows larger \cite{Xu14CTW}. In this paper, we focus on the design of downlink precoding for C-RANs with FD-MIMO RUs by accounting for the impact of fronthaul capacity limitations. Previous works \cite{Simeone09EURADVSP, Marsch09GLOBECOM, Yu14ITA, Park14SPMAG, Kang14arXiv} on precoding design for the downlink of C-RAN systems either assume fixed channel matrices with full channel state information (CSI), see \cite{Simeone09EURADVSP, Marsch09GLOBECOM, Yu14ITA, Park14SPMAG}, or considers ergodic channels with generic correlation structure and possibly imperfect CSI \cite{Kang14arXiv}. Importantly, these works do not account for the special features of FD channel models \cite{Ying14ICC, Alkhateeb14ASILOMAR} and hence do not bring insights into the feasibility of a C-RAN deployment based on FD-MIMO. In particular, the FD-MIMO channel is understood to be characterized by time variability at different time scales for elevation and azimuth components; elevation component changes significantly more slowly than the rate of change of the more conventional azimuth component \cite{Ying14ICC}. 

In order to address the design and performance of C-RAN system with FD-MIMO, this paper puts forth the following contributions.
\begin{itemize}
\item A \textit{layered precoding} scheme is proposed whereby separate precoding matrices are applied for the azimuth and elevation channel components with a different rate of adaptation to the channel variations. Specifically, a single precoding matrix is designed for the elevation channel across all coherence times based on stochastic CSI, while precoding matrices are optimized for the azimuth channel by adapting instantaneous CSI. This layered approach, considered in\cite{Alkhateeb14ASILOMAR} for a conventional cellular architecture, has the unique advantage in a C-RAN of potentially reducing the fronthaul transmission rate, due to the opportunity to amortize the overhead related to the elevation channel component across multiple coherence times.
\item We study layered precoding in a C-RAN system by considering two different CU-RU functional splits at the physical layer, namely the conventional C-RAN implementation, referred to as Compress-After-Precoding (CAP) as in \cite{Simeone09EURADVSP, Marsch09GLOBECOM, Yu14ITA, Park14SPMAG, Kang14arXiv}, whereby all baseband processing is done at the CU, and an alternative split, known as Compress-Before-Precoding (CBP) \cite{Kang14arXiv, Chae13ICC}, in which channel encoding and precoding are instead performed at the RUs. 
\item We carry out a performance comparison between standard non-layered precoding strategies and layered precoding for C-RAN systems with FD-MIMO under different functional splits as a function of system parameters such as the fronthaul capacity and the duration of the coherence period.
\end{itemize}

The rest of the paper is organized as follows. We describe the system model in Section II. In Section III, we review the conventional non-layered precoding schemes corresponding to the mentioned functional splits, namely CAP and CBP \cite{Kang14arXiv}. Then, we propose and optimize the layered precoding strategy for fronthaul compression in Section IV. In Section V, numerical results are presented. Concluding remarks are summarized in Section VI.

\emph{Notation}: $E[ \cdot ]$ and $\textrm{tr}( \cdot )$ denote the expectation and trace of the argument matrix, respectively. We use the standard notation for mutual information \cite{GamalBook}. ${\bf{\nu}}_{\textrm{max}} ({\bf{A}})$ is the eigenvector corresponding to the largest eigenvalue of the semi-positive definite matrix ${\bf{A}}$. We reserve the superscript ${\bf{A}}^{T}$ for the transpose of ${\bf{A}}$,  ${\bf{A}}^{\dagger}$ for the conjugate transpose of ${\bf{A}}$, and ${\bf{A}}^{-1} = ({\bf{A}}^\dagger {\bf{A}})^{-1} {\bf{A}}^\dagger$, which reduces to the usual inverse if the number of columns and rows are same. The identity matrix is denoted as ${\bf{I}}$. ${\bf{A}} \otimes {\bf{B}}$ is the Kronecker product of ${\bf{A}}$ and ${\bf{B}}$. 

\section{System Model} \label{Sec:SM}
We consider the downlink of a C-RAN in which a cluster of $N_R$ RUs provides wireless service to $N_M$ mobile stations (MSs) as illustrated in Fig. \ref{fig:fig1}. Each RU $i$ has a FD, or two-dimensional (2D), antenna array of $N_{A,i}$ horizontal antennas by $N_{E,i}$ vertical antennas and each MS has a single antenna. RU $i$ is connected to the CU via fronthaul link of capacity $\bar C_i$ bit per downlink symbol, where the downlink symbol rate equals the baud rate, i.e., no oversampling is performed.

\vspace{-0.4cm}
\subsection{Signal Model}
Each coded transmission block spans multiple coherence periods, e.g., multiple distinct resource blocks in an LTE system, of the downlink channel that contain $T$ symbols each. The $T \times 1$ signal ${\bf{y}}_j$ received by the MS $j$ in a given coherence interval is given by
\begin{equation} \label{RS;MS}
{\bf{y}}_{j} = {\bf{X}}^T {\bf{h}}_j  + {\bf{z}}_{j},
\end{equation}
where ${\bf{z}}_{j}$ is the $T \times 1$ noise vector with i.i.d. $\mathcal{CN}(0,1)$ components; ${\bf{h}}_j = [{\bf{h}}_{j 1}^T, \dots, {\bf{h}}_{j N_R}^T]^T$ denotes the $\sum_{i=1}^{N_R} N_{A,i} N_{E,i} \times 1$ channel vector for MS $j$, where ${\bf{h}}_{ji}$ is the $N_{A,i} N_{E,i} \times 1$ channel vector from the $i$-th RU to the MS $j$ as further discussed below; and $ {\bf{X}}$ is an $\sum_{i=1}^{N_R} N_{A,i} N_{E,i} \times T$ matrix that stacks the signals transmitted by all the RUs, i.e., $ {\bf{X}} = [ {\bf{X}}_{1}^T, \dots,  {\bf{X}}_{N_R}^T]^T$, where ${\bf{X}}_i$ is a $N_{A,i} N_{E,i} \times T$ complex baseband signal matrix transmitted by the $i$-th RU with each channel coherence period of duration $T$ channel uses. Note that each column of the signal matrix ${\bf{X}}_i$ corresponds to the signal transmitted from the $N_{A,i} N_{E,i}$ antennas in a channel use. The transmit signal ${\bf{X}}_i$ has a power constraint given as $E [ | {\bf{X}}_i |^2 ] = T \bar P_i$.

The channel vector ${\bf{h}}_j$ is assumed to be constant during each channel coherence block and to change according to a stationary ergodic process from block to block. We assume that the CU has perfect instantaneous information about the channel matrix ${\bf{H}} = [{\bf{h}}_1, \dots, {\bf{h}}_{N_M}]$ and MSs have full CSI about their respective channel matrices.
\vspace{-0.4cm}
\subsection{FD Channel Model} \label{Sec:Ch_Model}
\begin{figure}[t]
\centering
\vspace{-0.5cm}
\includegraphics[width=8cm]{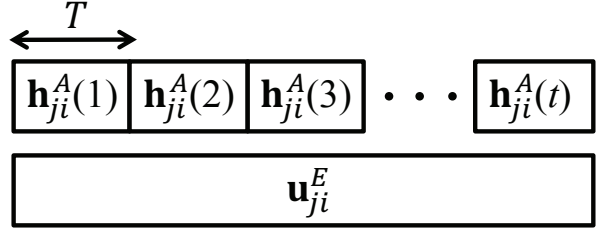}
\vspace{-0.5cm}
\caption{Illustration of time variability of the azimuth component $\{{\bf{h}}_{ji}^A (t)\}$ and of the elevation component ${\bf{u}}_{ji}^E$ in the FD channel model (\ref{CH_Model}). The notation ${\bf{h}}_{ji}^A (t)$ emphasizes the dependence on the coherence block $t$ of the azimuth component of the channel.}
\vspace{-0.5cm}
\label{fig:CHmodel}
\end{figure}

As in, e.g., \cite{Ying14ICC, Alkhateeb14ASILOMAR}, we assume that each RU is equipped with a uniform rectangular array (URA). Furthermore, the channel vector ${\bf{h}}_{ji}$ from RU $i$ to MS $j$ is modeled by means of a Kronecker product spatial correlation model \cite{Ying14ICC, Alkhateeb14ASILOMAR}. This was shown to provide a good modeling choice under the condition that the MS is sufficiently far away from the RUs \cite{Ying14ICC}. According to this model, the covariance of the 3D channel ${\bf{h}}_{ji}$ which is defined as ${\bf{R}}_{ji} =  E[{\bf{h}}_{ji} {\bf{h}}_{ji}^\dagger] $, is written as 
\begin{equation}
{\bf{R}}_{ji} = {\bf{R}}_{ji}^A \otimes {\bf{R}}_{ji}^E,
\end{equation}
where ${\bf{R}}_{ji}^A$ and ${\bf{R}}_{ji}^E$ represent the covariance matrices in the azimuth and elevation directions, respectively. Since the elevation direction is typically subject to negligible scattering \cite{Seifi14TWC, Zhong13VTC}, the elevation covariance matrix ${\bf{R}}_{ji}^E$ may be assumed to be a rank-1 matrix, i.e., ${\bf{R}}_{ji}^E = {\bf{u}}^E_{ji} {\bf{u}}_{ji}^{E \, \dagger}$, where ${\bf{u}}^E_{ji}$ is a $N_{E,i} \times 1$ unit-norm vector \cite{Alkhateeb14ASILOMAR}. Under this assumption, the channel vector ${\bf{h}}_{ji}$ can be written as 
\begin{equation} \label{CH_Model}
{\bf{h}}_{ji} = \sqrt{\alpha_{ji}} {\bf{h}}_{ji}^A \otimes {\bf{u}}_{ji}^E,
\end{equation}
where $\alpha_{ji}$ denotes the path loss coefficient between MS $j$ and RU $i$ as 
\begin{equation} \label{PL_coef}
\alpha_{ji} = \frac{1}{1 + \left (\frac{d_{ji}}{d_0}\right )^{\eta}},
\end{equation}
with $d_{ji}$ being the distance between the $j$-th MS and the $i$-th RU, $d_0$ being a reference distance, and $\eta$ being the path loss exponent; and ${\bf{h}}_{ji}^A \sim \mathcal{CN} (0, {\bf{R}}_{ji}^A)$ with ${\bf{R}}_{ji}^A$ having diagonal elements equal to one. This model entails that the elevation components ${\bf{h}}_{ji}^E$ remains constant over coherence interval, while the azimuth component changes independent across coherence interval as ${\bf{h}}_{ji}^A \sim \mathcal{CN}(0,{\bf{R}}_{ji}^A)$, as illustrated in Fig. \ref{fig:CHmodel}.

\vspace{-0.5cm}
\section{Background}
In this section, we briefly recall in an informal fashion two baseline strategies for downlink transmission in the C-RAN system introduced above. The strategies correspond to two different functional splits at the physical layer between CU and RUs \cite{Dotsch13Bell, Wubben14SPMAG} as detailed in \cite{Kang14arXiv}. We note that these schemes were previously proposed and studied without specific reference to FD-MIMO and hence do not leverage the special structure of the channel model (\ref{CH_Model}).

\begin{figure}[t]
\centering
\includegraphics[height=5cm]{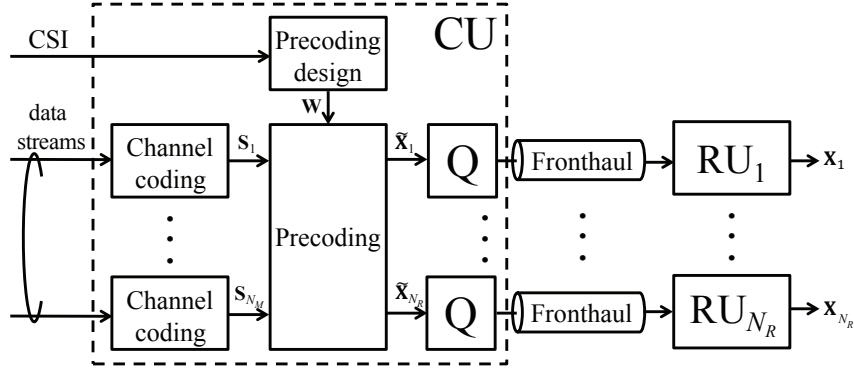}
\vspace{-.5cm}
\caption{Block diagram of the (non-layered) Compression-After-Precoding (CAP) scheme (``${\text{Q}}$" represents fronthaul compression).}
\vspace{-.5cm}
\label{fig:SM_CAP}
\end{figure}
\vspace{-0.3cm}
\subsection{Standard C-RAN Processing: Precoding at the CU} \label{CAP}
In the standard C-RAN approach, all baseband processing is done at the CU. Specifically, as illustrated in Fig. \ref{fig:SM_CAP}, the CU performs channel coding and precoding, and then compresses the resulting baseband signals so that they can be forwarded on the fronthaul links to the corresponding RUs. The RUs upconvert the received quantized baseband signal prior to transmission on the wireless channel. Following \cite{Kang14arXiv}, we refer to this strategy as Compression-After-Precoding (CAP). Analysis and optimization of the CAP strategy can be found in \cite{Kang14arXiv}.
\vspace{-0.3cm}
\subsection{Alternative Functional Split: Precoding at the RUs} \label{CBP}
\begin{figure}[t]
\centering
\includegraphics[height=5cm]{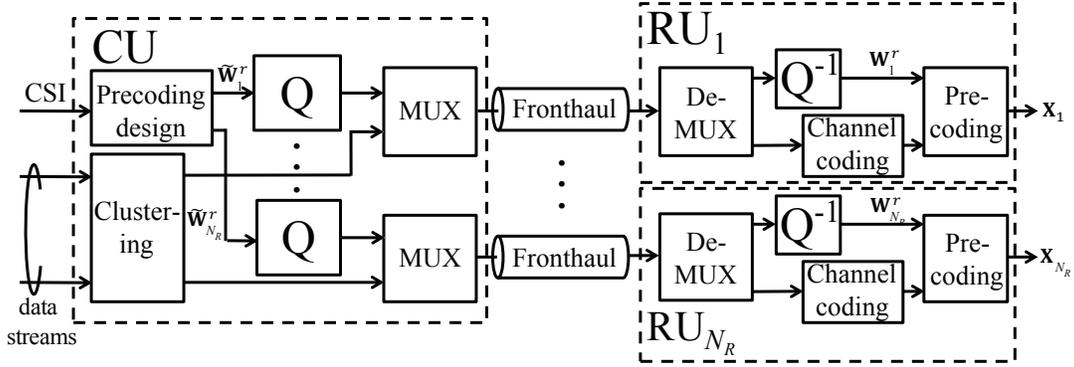}
\vspace{-.5cm}
\caption{Block diagram of the (non-layered) Compression-Before-Precoding (CBP) scheme (``${\text{Q}}$" represents fronthaul compression).}
\vspace{-.5cm}
\label{fig:SM_CBP}
\end{figure}

As an alternative to the standard C-RAN approach just described, one can instead implement channel encoding and precoding at the RUs. This is referred to as Compression-Before-Precoding (CBP) in \cite{Chae13ICC, Kang14arXiv}. According to this solution, as seen in Fig. \ref{fig:SM_CBP}, the CU calculates the precoding matrices based on the available CSI, but does not perform precoding. Instead, it uses the fronthaul links to communicate the downlink information streams to each RU, along with the compressed precoding matrix. Each RU can then encode and precode the messages of the MSs based on the information received from the fronthaul link. As elaborated on in \cite{Kang14arXiv}, this alternative functional split is generally advantageous when the number of MSs is not too large and when the coherence period $T$ is large enough. This is because, when the number of MSs is small, a lower fronthaul overhead is needed to communicate the data streams of the MSs on the fronthaul link; and, when the coherence period $T$ is large, the compressed precoding information can be amortized over a longer period, hence reducing the fronthaul rate.

\section{Layered Precoding for Reduced Fronthaul Overhead} \label{Hybrid_Strategy}
The baseline state-of-the-art fronthaul transmission strategies mentioned above do not make any provision to exploit the special structure of the FD channel model (\ref{CH_Model}), and can hence be inefficient if the number of vertical antennas is large. In this section, we propose a layered precoding that instead leverages the different dynamic characteristic of the elevation and azimuth channels as per channel model (\ref{CH_Model}). We recall that, according to this model, the elevation channel has a constant direction across the coherence periods in its elevation component due to the rank-1 covariance matrix, while its azimuth component changes in each coherence period due to the generally larger rank of its covariance matrix (see Fig. \ref{fig:CHmodel}).

In order to exploit this channel decomposition, we propose that the CU designs separate precoding matrices for the elevation and azimuth channels following a layered precoding approach. The key idea is that of designing a single precoding matrix for the elevation channel across all coherence times based on long-term CSI, while adapting only the azimuth precoding matrix to the instantaneous channel conditions. This allows the CU to accurately describe the elevation precoding matrix through the fronthaul links via quantization with negligible overhead given that the latter is amortized across all coherence periods. Precoding on the azimuth channel can instead be handled via either a CAP or CBP-like scheme, as detailed below. 

In the following, we first describe the layered precoding approach in Section \ref{LP:HS}; then introduce the precoding and fronthaul compression strategy based on CAP in Section \ref{PFC:HS}; and, finally, we introduce CBP-based fronthaul compression and layered precoding design in Section \ref{CBP_FC:HS}. 

\begin{figure}[t]
\centering
\vspace{-0.5cm}
\includegraphics[width=8cm]{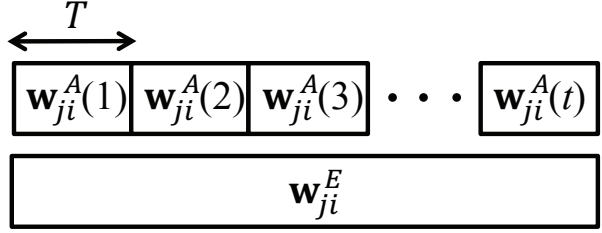}
\vspace{-0.5cm}
\caption{Illustration of time variability of the azimuth and elevation components of beamforming in the layered precoding scheme (\ref{LP_Model}).}
\vspace{-0.5cm}
\label{fig:LPmodel}
\end{figure}
\subsection{Layered Precoding} \label{LP:HS}
Leveraging the channel decomposition resulting from the Kronecker channel model (\ref{CH_Model}), we propose to factorize the $N_{t,i} \times 1$ precoding vector ${\bf{w}}_{ji}$ for RU $i$ toward MS $j$ as 
\begin{equation} \label{LP_Model}
{\bf{w}}_{ji} = {\bf{w}}_{ji}^A \otimes {\bf{w}}_{ji}^E,
\end{equation}
where ${\bf{w}}_{ji}^A$ denotes the $N_{A,i} \times 1$ azimuth component and ${\bf{w}}_{ji}^E$ is the $N_{E,i} \times 1$ elevation component of the precoding vector for MS $j$ and RU $i$ designed based on the elevation channels. A similar model was proposed in \cite{Alkhateeb14ASILOMAR} for co-located antenna arrays. The corresponding $N_{A,i} \times N_M$ azimuth precoding matrix ${\bf{W}}_i^A$ and the $N_{E,i} \times N_M$ elevation precoding matrix ${\bf{W}}_i^E$ for RU $i$ are defined as ${\bf{W}}_i^A = [{\bf{w}}_{1i}^A, \dots, {\bf{w}}_{N_M i}^A]$ and ${\bf{W}}_i^E = [{\bf{w}}_{1i}^E, \dots, {\bf{w}}_{N_M i}^E]$, respectively. In the proposed solutions, each elevation component ${\bf{w}}_{ji}^E$ is quantized by the CU and sent to the $j$-th RU via the corresponding fronthaul links. Since this vector is to be used for all coherence times, as illustrated in Fig. \ref{fig:LPmodel}, its fronthaul overhead can be amortized across multiple coherence interval. As a result, it can be assumed to be known accurately at the RUs. Moreover, the corresponding fronthaul overhead for the transfer of elevation precoding information on the fronthaul links can be assumed to be negligible. For the azimuth components, we may adopt either a CAP or CBP approach, as discussed next. 
\subsection{CAP-based Fronthaul Compression for Layered Precoding} \label{PFC:HS}
\begin{figure}[t]
\centering
\includegraphics[height=6.3cm]{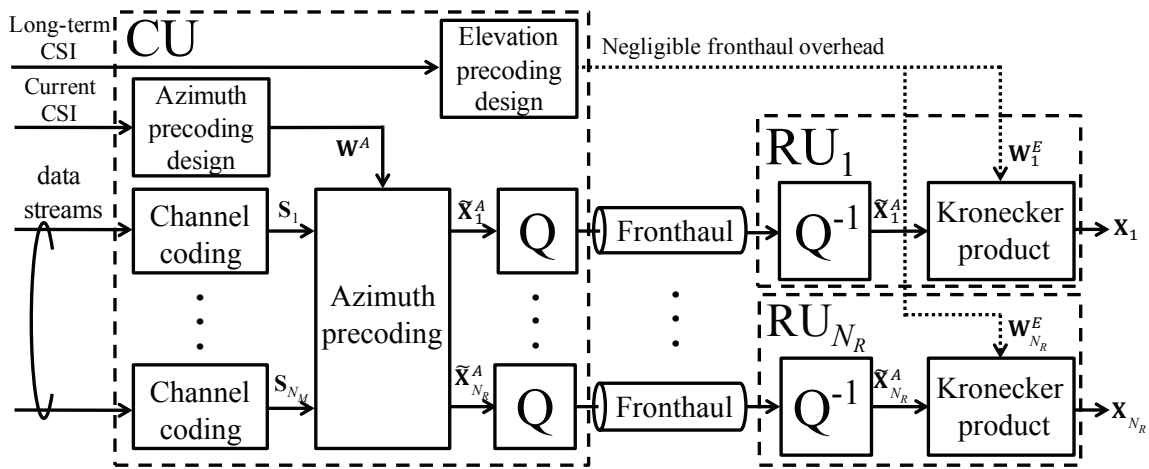}
\vspace{-.5cm}
\caption{Block diagram of the Layered Compression-After-Precoding (CAP) scheme (``${\text{Q}}$" represents fronthaul compression).}
\vspace{-.5cm}
\label{fig:fig2}
\end{figure}

In the proposed  CAP-based solution, the CU applies precoding only for the azimuth component. Accordingly, the azimuth-precoded baseband signals, as well as the precoding matrix for the elevation component, are separately compressed at the CU and forwarded over the fronthaul links to each RU. In order to perform precoding over both elevation and azimuth channels, each RU finally performs the Kronecker product of the compressed baseband signal ${\bf{X}}_{ji}^A$ and the precoding vector ${\bf{w}}_{ji}^E$ for elevation channel. A block diagram can be found in Fig. \ref{fig:fig2} and details are provided next.

\subsubsection{Details and Analysis}
Let $\widetilde {\bf{X}}_{ji}^A$ be the $N_{A,i} \times T$ precoded signal only for the azimuth channel between RU $i$ and MS $j$ in a given coherence period. This is defined as $\widetilde {\bf{X}}_{ji}^A = {\bf{w}}_{ji}^A {\bf{s}}_j^T$, where ${\bf{s}}_j$ is the $T \times 1$ vector containing the encoded data stream for MS $j$ in the given coherence period. Note that all the entries of vector ${\bf{s}}_{j}$ are assumed to have i.i.d. $\mathcal{CN} (0,1)$ from standard random coding arguments. Adopting a CAP-like approach, the CU quantizes each sequence of baseband signals $\{ \widetilde {\bf{X}}_{ji}^A \}$, for all $j \in \mathcal{N}_M$, across all coherence periods intended for RU $i$ for transfer on $i$-th fronthaul. The compressed signal ${\bf{X}}_{ji}^A$ is modeled as
\begin{equation} \label{Compress;LCAP}
{\bf{X}}_{ji}^A = \widetilde {\bf{X}}_{ji}^A + {\bf{Q}}_{x,ji}^A = {\bf{w}}_{ji}^A {\bf{s}}_j^T + {\bf{Q}}_{x,ji}^A,
\end{equation}
where ${\bf{Q}}_{x,ji}^A$ is the quantization noise matrix, which is assumed to have i.i.d. $\mathcal{CN} (0, \sigma_{x,ji}^{2})$ entries. From standard rate-distortion arguments \cite{GamalBook, CoverBook}, the required rate for transfer of the precoded data signals $\{ \widetilde {\bf{X}}^A_{ji} \}_{j \in \mathcal{N}_M}$ on fronthaul link between the CU and RU $i$ is given as 
\begin{equation} \label{FCConstraint;CAP}
C_{x,i} ({\bf{W}}_i^A, {\pmb{\sigma}}_{x,i}^2)  = \sum_{j=1}^{N_M} I \left( {\bf{X}}_{ji}^A; \widetilde {\bf{X}}_{ji}^A \right) = \sum_{j=1}^{N_M} \left \{ \log \left( ||{\bf{w}}_{ji}^A||^2 + \sigma_{x,ji}^{2} \right) - \log \sigma_{x,ji}^{2} \right \},
\end{equation}
where we have used the assumption that the data signal ${\bf{X}}_{ji}^A$ are independent across the MS index $j$ and we have defined ${\pmb{\sigma}}_{x,i}^2 = [{{\sigma}}_{x,1i}^2, \dots, {{\sigma}}_{x,N_M i}^2 ]^T$. Note that, unlike the standard CAP scheme, here the signals for different MSs are separately compressed as per (\ref{Compress;LCAP}).

Considering also the elevation component, the resulting signal ${\bf{X}}_{i}$ computed and transmitted by RU $i$ is obtained as ${\bf{X}}_i = \sum_{j=1}^{N_M} {\bf{X}}_{ji}$, with 
\begin{equation}
{\bf{X}}_{ji} = {\bf{X}}_{ji}^A \otimes {\bf{w}}_{ji}^E = ({\bf{w}}_{ji}^A {\bf{s}}_j^T + {\bf{Q}}_{x,ji}^A) \otimes {\bf{w}}_{ji}^E = ({\bf{w}}_{ji}^A \otimes {\bf{w}}_{ji}^E) {\bf{s}}_j^T + {\bf{Q}}_{x,ji}^A \otimes {\bf{w}}_{ji}^E. 
\end{equation}
The power transmitted at RU $i$ is then computed as
\begin{eqnarray} \label{PowerConst}
P_i ({\bf{W}}_i^A, {\bf{W}}_i^E, {\pmb{\sigma}}_{x,i}^2) = \textrm{tr} \left ({\bf{X}}_{i}{\bf{X}}_{i}^\dagger \right) &=& \textrm{tr} \left (\sum_{j=1}^{N_M} \left( \left( {\bf{w}}_{ji}^A {\bf{s}}_j^T + {\bf{Q}}^A_{x,ji} \right) \otimes {\bf{w}}_{ji}^E \right) \left( \left( {\bf{w}}_{ji}^A {\bf{s}}_j^T + {\bf{Q}}_{x,ji}^A \right) \otimes {\bf{w}}_{ji}^E \right)^\dagger \right) \\
\nonumber &=& \sum_{j=1}^{N_M} \left ( ||{\bf{w}}_{ji}^A||^2 ||{\bf{w}}_{ji}^E||^2 +  N_{A,i} \sigma_{x,ji}^2 ||{\bf{w}}_{ji}^E||^2 \right ),
\end{eqnarray}
where we have used the property of the Kronecker product that $({\bf{A}} \otimes {\bf{B}}) ({\bf{C}} \otimes {\bf{D}}) = ({\bf{A}}{\bf{C}} \otimes {\bf{B}} {\bf{D}})$ and ${\textrm{tr}} ({\bf{A}} \otimes {\bf{B}}) =  {\textrm{tr}} ( {\bf{A}}) {\textrm{tr}} ({\bf{B}})$ \cite{BrookesOnline}. 

The ergodic achievable rate for MS $j$ is evaluated as $E [ R_j ({\bf{H}}, {\bf{W}}^A, {\bf{W}}^E, {\pmb{\sigma}}_{x}^2)]$, with $R_j ({\bf{H}}, {\bf{W}}^A, {\bf{W}}^E, {\pmb{\sigma}}_{x}^2) = I_{\bf{H}} ({\bf{s}}_j; {\bf{y}}_j) /T$, where $I_{\bf{H}} ({\bf{s}}_j; {\bf{y}}_j)$ is the mutual information conditioned on the value of channel matrix ${\bf{H}}$, the expectation is taken with respect to ${\bf{H}}$ and
\begin{eqnarray} \label{ASR_HS}
&& \hspace{-1cm} R_j ({\bf{H}}, {\bf{W}}^A, {\bf{W}}^E, {\pmb{\sigma}}_{x}^2) = \log \left( 1 + \sum_{k=1}^{N_M} \sum_{i=1}^{N_R} \lambda_{ji}^E | {\bf{u}}_{ji}^E {\bf{w}}_{ki}^E |^2 \left( | {\bf{w}}_{ki}^{A \, \dagger}  {\bf{h}}_{ji}^{A} |^2 + \sigma_{x,ki}^2 ||{\bf{h}}_{ji}^{A}||^2  \right ) \right ) \\
\nonumber&& \hspace{5cm} - \log \left ( 1 + \sum_{k=1,k \neq j}^{N_M} \sum_{i=1}^{N_R} \lambda_{ji}^E | {\bf{u}}_{ji}^E {\bf{w}}_{ki}^E |^2 \left( | {\bf{w}}_{ki}^{A \, \dagger}  {\bf{h}}_{ji}^{A} |^2 + \sigma_{x,ki}^2 ||{\bf{h}}_{ji}^{A}||^2  \right ) \right ),
\end{eqnarray}
where ${\bf{W}}^A = [({\bf{W}}^{A}_1)^T, \dots, ({\bf{W}}^{A}_{N_R})^T]^T$, ${\bf{W}}^E = [({\bf{W}}^{E}_1)^T, \dots, ({\bf{W}}^{E}_{N_R})^T]^T$, and ${\pmb{\sigma}}_x^2 = [{\pmb{\sigma}}_{x,1}^2, \dots, {\pmb{\sigma}}_{x,N_R}^2 ]$.
\linespread{1.5}
\begin{algorithm} [t]
\begin{algorithmic}
\caption{CAP-based Fronthaul Compression and Layered Precoding Design} \label{Algorithm_HS}
\State {\textbf{1) Long-term Optimization of Elevation Precoding}}
\State {\textbf{Input:}} Long-term statistics of the channel
\State {\textbf{Output:}} Elevation precoding ${{\bf{W}}^{E}}^{{\pmb{*}}}$
\State {\textbf{Initialization (outer loop)}}: Initialize the covariance matrix ${\bf{V}}^{E \, (n)} \succeq 0$ subject to ${\textrm{tr}} ({\bf{V}}^{E \, (n)}) = 1$ and set $n=0$. 
\State  {\textbf{Repeat}}  
\State \indent $n \gets n+1$
\State \indent Generate a channel matrix realization ${\bf{H}}^{(n)}$ using the available stochastic CSI.
\State \indent {\textbf{Inner loop}}: Obtain ${\bf{V}}^{\hspace{-0.05cm} A (n)}\hspace{-0.05cm}(\hspace{-0.05cm}{\bf{H}}^{(n)}\hspace{-0.05cm})$ and ${\pmb{\sigma}}_{x}^{2 (n)}\hspace{-0.05cm}(\hspace{-0.05cm}{\bf{H}}^{(n)}\hspace{-0.05cm})$ with ${\bf{V}}^{\hspace{-0.05cm} E} \gets {\bf{V}}^{\hspace{-0.05cm} E (n-1)}$ using Algorithm \ref{Algorithm_DC}.
\State \indent Update ${\bf{V}}^{E \, (n)}$ by solving problem (\ref{OP_wSSUM_HS}), which depends on ${\bf{V}}^{A \, (m)}({\bf{H}}^{(m)})$ and ${\pmb{\sigma}}_{x}^{2 \, (m)}({\bf{H}}^{(m)})$
\State \indent for all $m \le n$.
\State {\textbf{Until}} a convergence criterion is satisfied.
\State Set ${\bf{V}}^{E} \gets {\bf{V}}^{E \, (n)}$.
\State {\textbf{Calculation of} ${{\bf{W}}^{E}}^{{\pmb{*}}}$}: Calculate the precoding matrix ${{\bf{W}}^E}^{{\pmb{*}}}$ for elevation channel from the covariance matrix ${\bf{V}}^{E}$ via rank reduction as ${{\bf{w}}^E_{ji}}^{{\pmb{*}}} = {\mathbf{\nu}}_{\textrm{max}} ({\bf{V}}_{ji}^{E})$ for all $j \in \mathcal{N}_M$ and $i \in \mathcal{N}_R$.
\State {\textbf{2) Short-term Optimization of Azimuth Precoding and Quantization Noise }}
\State {\textbf{Input:}} Channel ${\bf{H}}$ and elevation precoding ${{\bf{W}}^E}^{{\pmb{*}}}$
\State {\textbf{Output:}} Azimuth precoding ${{\bf{W}}^{A}}^{{\pmb{*}}} ({\bf{H}})$ and quantization noise vector ${{\pmb{\sigma}}_x^2}^{{\pmb{*}}} ({\bf{H}})$
\State Obtain ${\bf{V}}^{A}({\bf{H}})$ and ${\pmb{\sigma}}_{x}^{2} ({\bf{H}})$ with ${\bf{W}}^E \gets {{\bf{W}}^{E}}^{{\pmb{*}}}$ using Algorithm \ref{Algorithm_DC}.
\State {\textbf{Calculation of} ${{\bf{W}}^A}^{{\pmb{*}}} ({\bf{H}})$}: Calculate the precoding matrix ${{\bf{W}}^{A}}^{{\pmb{*}}} ({\bf{H}})$ for the azimuth channel from the covariance matrix ${\bf{V}}^{A}({\bf{H}})$ via rank reduction as ${{\bf{w}}^A_{ji}}^{{\pmb{*}}} ({\bf{H}}) = \beta_{ji} {\mathbf{\nu}}_{\textrm{max}} ({\bf{V}}_{ji}^{A}({\bf{H}}))$ for all $j \in \mathcal{N}_M$ and $i \in \mathcal{N}_R$, where $\beta_{ji}$ is obtained by imposing $P_i ({{\bf{W}}_i^A}^{{\pmb{*}}} ({\bf{H}}), {{\bf{W}}_i^E}^{{\pmb{*}}}, {{\pmb{\sigma}}_{x,i}^2}^{{\pmb{*}}} ({\bf{H}})) = \bar P_i$ using (\ref{PowerConst}).
\end{algorithmic}
\end{algorithm}
\linespread{2}
\linespread{1.5}
\begin{algorithm} [t]
\begin{algorithmic}
\caption{DC Algorithm for Optimization of ${\bf{V}}^A({\bf{H}})$ and ${\pmb{\sigma}}_{x}^{2}({\bf{H}})$} \label{Algorithm_DC}
\State {\textbf{Input:}} Channel ${\bf{H}}$ and elevation precoding ${\bf{V}}^E$.
\State {\textbf{Output:}} ${\bf{V}}^A({\bf{H}})$ and ${\pmb{\sigma}}_x^2({\bf{H}})$
\State {\textbf{Initialization}}: Initialize ${\bf{V}}^{A \, (0)}({\bf{H}}) \succeq 0$ and ${\pmb{\sigma}}_{x}^{2 \, (0)}({\bf{H}}) \in {\mathbb{R}}^+$, and set $l=0$. 
\State  {\textbf{Repeat}}  
\State \indent $l \gets l+1$ 
\State \indent Update ${\bf{V}}^{A \, (l)}({\bf{H}})$ and ${\pmb{\sigma}}_{x}^{2 \, (l)}({\bf{H}})$ by solving problem (\ref{OP_wMM_HS}).
\State {\textbf{Until}} a convergence criterion is satisfied.
\State Set ${\bf{V}}^{A}({\bf{H}}) \gets {\bf{V}}^{A \, (l)}({\bf{H}})$ and ${\pmb{\sigma}}_{x}^{2}({\bf{H}}) \gets {\pmb{\sigma}}_{x}^{2 \,\, (l)}({\bf{H}})$.
\end{algorithmic}
\end{algorithm}
\linespread{2}
\subsubsection{Problem Formulation}
The ergodic achievable sum-rate (\ref{ASR_HS}) can be optimized over the precoding matrices ${\bf{W}}^A$ and ${\bf{W}}^E$, and over the quantization noise variance vector ${\pmb{\sigma}}_{x}^2$ under fronthaul capacity and power constraints. Since the design of the precoding matrix ${\bf{W}}^A$ for azimuth channel and of the compression noise variance ${\pmb{\sigma}}_{x}^2$ is adapted to the channel realization ${\bf{H}}$ for each coherence block, we use the notations ${\bf{W}}^A({\bf{H}})$ and ${\pmb{\sigma}}_{x}^2 ({\bf{H}})$. The problem of maximizing the achievable rate is then formulated as follows
\begin{subequations} \label{OP_HS}
\begin{eqnarray} 
\underset { {\bf{W}}^A({\bf{H}}), {\bf{W}}^E, {\pmb{\sigma}}_{x}^2({\bf{H}})}{\textrm{maximize}} && \sum_{j \in \mathcal{N}_M} E [ R_j ({\bf{H}}, {\bf{W}}^A({\bf{H}}), {\bf{W}}^E, {\pmb{\sigma}}_{x}^2({\bf{H}})) ] \label{OF_OP;HS} \\
\textrm{s.t.} && C_{x,i} ({\bf{W}}_i^A({\bf{H}}), {\pmb{\sigma}}_{x,i}^2({\bf{H}})) \le \bar C_i, \hspace{1cm} \forall i \in \mathcal{N}_R, \label{FC_OP;HS} \\
&& P_i ({\bf{W}}_i^A({\bf{H}}), {\bf{W}}_i^E, {\pmb{\sigma}}_{x,i}^2({\bf{H}})) \le \bar P_i, \hspace{1cm} \forall i \in \mathcal{N}_R, \label{PC_OP;HS}
\end{eqnarray}
\end{subequations}
where the constraints apply for all channel realizations ${\bf{H}}$, and we recall that the capacity constraint on $i$-th fronthaul link is $\bar C_i$ and the power constraint for RU $i$ is $\bar P_i$. 
\subsubsection{Optimization Algorithm} In problem (\ref{OP_HS}), the objective function (\ref{OF_OP;HS}) and constraint (\ref{FC_OP;HS}) are non-convex in terms of ${\bf{W}}^A({\bf{H}})$, ${\bf{W}}^E$, and ${\pmb{\sigma}}_{x}^2 ({\bf{H}})$. Furthermore, as discussed above, ${\bf{W}}^E$ is designed based on stochastic CSI (long-term CSI), while ${\bf{W}}^A({\bf{H}})$ and ${\pmb{\sigma}}_{x}^2 ({\bf{H}})$ are adapted to instantaneous CSI (short-term CSI). In order to tackle this problem, we propose an algorithm that optimizes separately the long-term and short-term variables ${\bf{W}}^E$ and $({\bf{W}}^A({\bf{H}}), {\pmb{\sigma}}_x^2 ({\bf{H}}))$, respectively. For the former optimization, we adopt a stochastic optimization approach based empirical approximation of the ensemble averages in (\ref{OF_OP;HS}) following Stochastic Successive Upper-bound Minimization (SSUM) method \cite{SSUM_paper}. For the latter, we instead invoke the Difference of Convex (DC) method \cite{MMBook, MMBook_tutorial} by leveraging the rank relaxation in obtained by reformulating the optimization problem in terms of the covariance matrices ${\bf{V}}^A_{ji}({\bf{H}}) = {\bf{w}}_{ji}^A({\bf{H}}){\bf{w}}_{ji}^{A \, \dagger}({\bf{H}})$ and ${\bf{V}}^E_{ji} = {\bf{w}}^E_{ji}{\bf{w}}_{ji}^{E \, \dagger}$ for all $j \in \mathcal{N}_M$ and $i \in \mathcal{N}_R$. The resulting algorithm is detailed in Algorithm \ref{Algorithm_HS} and Appendix \ref{Apx;Opt_CAP}. Note that, in Algorithm \ref{Algorithm_HS}, long-term optimization has two nested loops in which inner loop requires at each iteration the solution of a convex problem, whose complexity is polynomial in the problem size \cite{BoydBook}. 
\subsection{CBP-based Fronthaul Compression for Layered Precoding} \label{CBP_FC:HS}
\begin{figure}[t]
\centering
\includegraphics[height=6.8cm]{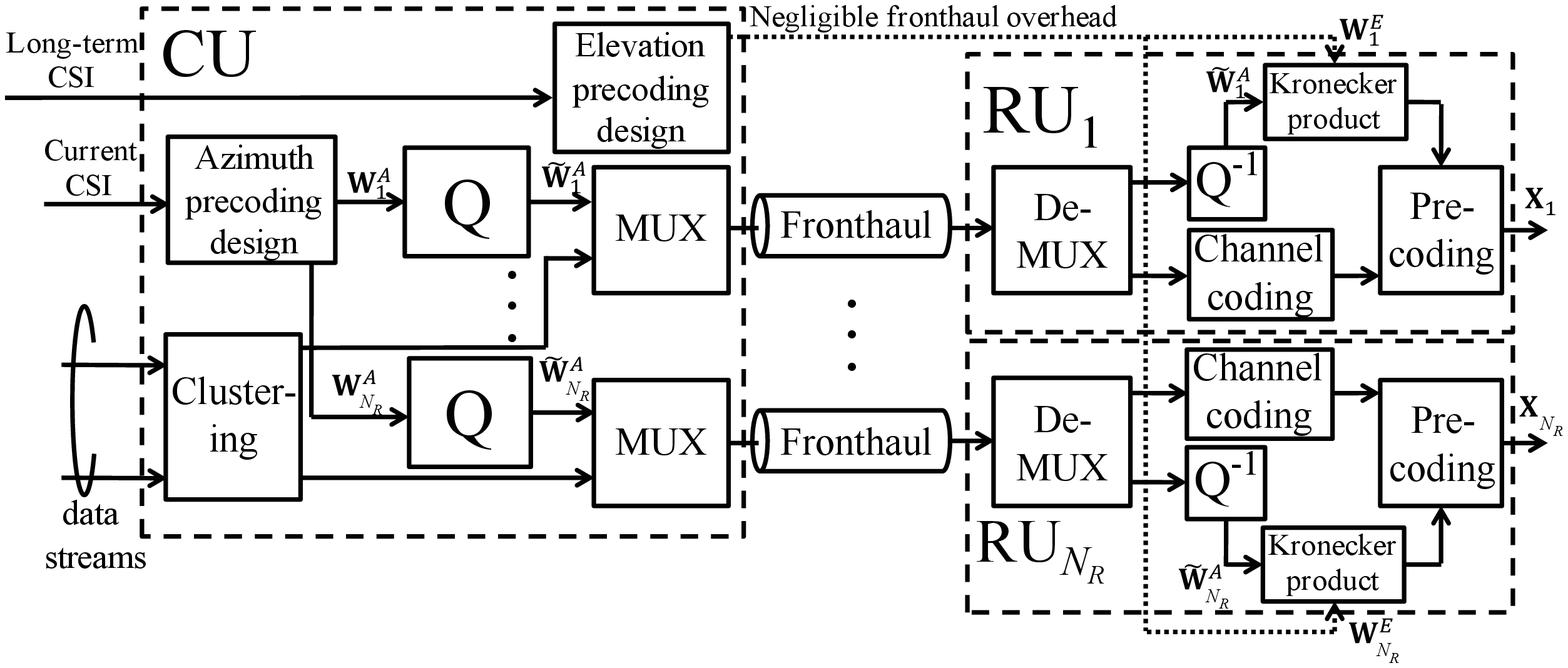}
\vspace{-.5cm}
\caption{Block diagram of the Layered Compression-Before-Precoding (CBP) scheme (``${\text{Q}}$" represents fronthaul compression).}
\vspace{-.5cm}
\label{fig:fig3}
\end{figure}

In the proposed CBP-based strategy, as illustrated in Fig. \ref{fig:fig3}, the CU designs the precoding matrices for both azimuth and elevation components, which are transferred, along with a given subset of downlink information messages, over the fronthaul link to the each RU. As discussed, since the design of the elevation precoding is done based on long-term CSI, and hence entails the use of a negligible portion of the fronthaul capacity, the fronthaul overhead depends only on the azimuth precoding matrices, which are adapted to current CSI, and on the information messages. As in \cite{Kang14arXiv}, the subset of information messages sent to each RU is determined by a preliminary clustering step at the CU whereby each RU is assigned to serve a subset of the MSs.  At each RU, the precoding matrix for FD-MIMO is computed via the Kronecker product between the precoding matrices for the azimuth and elevation channels. Based on the calculated precoding matrix, each RU can then encode and precode the received messages of the assigned MSs. Details are provided next.
\subsubsection{Details and Analysis}
To elaborate, let us denote the set of MSs assigned by RU $i$ as $\mathcal{M}_i \subseteq \mathcal{N}_M$, for all $i \in \mathcal{N}_R$. We also use $\mathcal{M}_i [k]$ to denote the $k$-th MS in the set $\mathcal{M}_i$. Note that we assume that the assignment of MSs is given and not subject to optimization. The azimuth precoding vectors $ \widetilde {\bf{W}}_{i}^A$ intended for RU $i$ are compressed by the CU and forwarded over the fronthaul link to RU $i$. The compressed azimuth precoding ${\bf{W}}_{i}^A$ for RU $i$ at the CU is then given by 
\begin{equation}
{\bf{W}}_{i}^A = \widetilde {\bf{W}}_{i}^A + {\bf{Q}}_{w,i},
\end{equation}
where the quantization noise matrix ${\bf{Q}}_{w,i}$ is assumed to have zero-mean i.i.d. $\mathcal{CN} (0, \sigma_{w,i}^2)$ entries. The required rate for the transfer of the azimuth precoding on fronthaul link is given, similar to (\ref{FCConstraint;CAP}), as 
\begin{eqnarray} \label{FC;LayeredCBP}
C_{w,i} (\widetilde {\bf{W}}_i^A, {{\sigma}}_{w,i}^2) &=& \frac{1}{T} I \left ( {\bf{W}}_{i}^A; \widetilde {\bf{W}}_{i}^A \right)\\
\nonumber &=& \frac{1}{T} \{ \log \det \left ( \widetilde {\bf{W}}_{i}^A \widetilde {\bf{W}}_{i}^{A \, \dagger} + \sigma_{w,i}^2 {\bf{I}} \right) - \log \det \left ( \sigma_{w,i}^2 {\bf{I}} \right) \},
\end{eqnarray}
where $\widetilde {\bf{W}}_i^A = [\widetilde {\bf{w}}_{\mathcal{M}_i [1] \, i}^A, \dots, \widetilde {\bf{w}}_{\mathcal{M}_i [|\mathcal{M}_i|] \, i}^A]$. The remaining fronthaul capacity is used to convey information messages, whose total rate is $\sum_{j \in \mathcal{M}_i} R_j$ with $R_j$ being the user rate for MS $j$. At each RU $i$, the precoding matrix for FD-MIMO is obtained via the Kronecker product of the elevation and azimuth components, yielding the transmitted signal ${\bf{X}}_i = \sum_{j \in \mathcal{M}_i} {\bf{X}}_{ji}$, with 
\begin{equation}
{\bf{X}}_{ji} = ( {\bf{w}}_{ji}^A \otimes {\bf{w}}_{ji}^E ) {\bf{s}}_j^T =  (\widetilde {\bf{w}}_{ji}^A \otimes {\bf{w}}_{ji}^E) {\bf{s}}_j^T + {\bf{q}}_{w,ji}^A {\bf{s}}_j^T \otimes {\bf{w}}_{ji}^E. 
\end{equation}
The power transmitted at RU $i$ is then calculated as
\begin{equation} \label{PowerConstraint;CBP}
P_i (\widetilde {\bf{W}}_i^A, {\bf{W}}_i^E, {{\sigma}}_{w,i}^2) = \textrm{tr} \left ({\bf{X}}_{i}{\bf{X}}_{i}^\dagger \right) =  \sum_{j \in \mathcal{M}_i} \left ( ||{\bf{w}}_{ji}^A||^2 ||{\bf{w}}_{ji}^E||^2 +  N_{A,i} \sigma_{w,i}^2 ||{\bf{w}}_{ji}^E||^2 \right ).
\end{equation}
\linespread{1.5}
\begin{algorithm} [t]
\begin{algorithmic}
\caption{CBP-based Fronthaul Compression and Layered Precoding Design} \label{Algorithm_CBP}
\State {\textbf{1) Long-term Optimization of Elevation Precoding and User Rates}}
\State {\textbf{Input:}} Long-term statistics of the channel and clustering $\{ \mathcal{M}_i \}$
\State {\textbf{Output:}} Elevation precoding ${\bf{W}}^{E \, *}$ and MSs' rates $\{R_j\}$
\State {\textbf{Initialization (outer loop)}}: Initialize the covariance matrix ${\bf{V}}^{E \, (n)} \succeq 0 $ subject to ${\textrm{tr}} ({\bf{V}}^{E \, (n)}) = 1$ and $\{ R_j^{(n)} \} \in {\mathbb{R}}^+$, and set $n=0$. 
\State  {\textbf{Repeat}}  
\State \indent $n \gets n+1$
\State \indent Generate a channel matrix realization ${\bf{H}}^{(n)}$ using the available stochastic CSI.
\State \indent {\textbf{Inner loop}}: Obtain $\widetilde {\bf{V}}^{\hspace{-0.05cm} A (n)}\hspace{-0.05cm}(\hspace{-0.05cm}{\bf{H}}^{(n)}\hspace{-0.05cm})$ and ${\pmb{\sigma}}_{w}^{2 (n)}\hspace{-0.05cm}(\hspace{-0.05cm}{\bf{H}}^{(n)}\hspace{-0.05cm})$ with ${\bf{V}}^{\hspace{-0.05cm} E} \gets {\bf{V}}^{\hspace{-0.05cm} E (n-1)}$ using Algorithm \ref{Algorithm_DC;CBP}.
\State \indent Update ${\bf{V}}^{E \, (n)}$ and $\{ R_j^{(n)} \}$ by solving problem (\ref{OP_wSSUM_CBP}), which depends on $\widetilde {\bf{V}}^{A \, (m)}({\bf{H}}^{(m)})$ and 
\State \indent ${\pmb{\sigma}}_{w}^{2 \, (m)}({\bf{H}}^{(m)})$ for all $m \le n$.
\State {\textbf{Until}} a convergence criterion is satisfied.
\State Set ${\bf{V}}^{E} \gets {\bf{V}}^{E \, (n)}$ and $\{ R_j \} \gets \{ R_j^{(n)} \}$.
\State {\textbf{Calculation of} ${{\bf{W}}^E}^{{\pmb{*}}}$}: Calculate the precoding matrix ${{\bf{W}}^E}^{{\pmb{*}}}$ for elevation channel from the covariance matrix ${\bf{V}}^{E}$ via rank reduction as ${{\bf{w}}^E_{ji}}^{{\pmb{*}}} = {\mathbf{\nu}}_{\textrm{max}} ({\bf{V}}_{ji}^{E})$ for all $j \in \mathcal{N}_M$ and $i \in \mathcal{N}_R$.
\State {\textbf{2) Short-term Optimization of Azimuth Precoding and Quantization Noise }}
\State {\textbf{Input:}} Channel ${\bf{H}}$ and elevation precoding ${{\bf{W}}^E}^{{\pmb{*}}}$
\State {\textbf{Output:}} Azimuth precoding $\widetilde {\bf{W}}^{A \, {{\pmb{*}}}} ({\bf{H}})$ and quantization noise vector ${{\pmb{\sigma}}_w^2}^{{\pmb{*}}} ({\bf{H}})$
\State Obtain $\widetilde {\bf{V}}^{A}({\bf{H}})$ and ${\pmb{\sigma}}_{w}^{2} ({\bf{H}})$ with ${\bf{W}}^E \gets {{\bf{W}}^{E}}^{{\pmb{*}}}$ using Algorithm \ref{Algorithm_DC;CBP}.
\State {\textbf{Calculation of} $\widetilde {\bf{W}}^{A \, {{\pmb{*}}}} ({\bf{H}})$}: Calculate the precoding matrix $\widetilde {\bf{W}}^{A \, {{\pmb{*}}}} ({\bf{H}})$ for the azimuth channel from the covariance matrix $\widetilde {\bf{V}}^{A}({\bf{H}})$ via rank reduction as $\widetilde {\bf{w}}^{A \, {{\pmb{*}}}}_{ji} ({\bf{H}}) = \beta_{ji} {\mathbf{\nu}}_{\textrm{max}} (\widetilde {\bf{V}}_{ji}^{A}({\bf{H}}))$ for all $j \in \mathcal{N}_M$ and $i \in \mathcal{N}_R$, where $\beta_{ji}$ is obtained by imposing $P_i (\widetilde {\bf{W}}_i^{A \, {{\pmb{*}}}} ({\bf{H}}), {\bf{W}}_i^E, {{{\sigma}}_{w,i}^2}^{{\pmb{*}}} ({\bf{H}}) ) = \bar P_i$ using (\ref{PowerConstraint;CBP}).
\end{algorithmic}
\end{algorithm}
\linespread{2}
The ergodic achievable rate for MS $j$ is calculated as $E [ \bar R_j ({\bf{H}}, \widetilde {\bf{W}}^A, {\bf{W}}^E, {\pmb{\sigma}}_{w}^2 )]$ with
\begin{eqnarray}
&& \hspace{-1cm} \bar R_j ({\bf{H}}, \widetilde {\bf{W}}^A, {\bf{W}}^E, {\pmb{\sigma}}_{w}^2) = \log \left( 1 +  \sum_{i=1}^{N_R} \sum_{k \in \mathcal{M}_i} \lambda_{ji}^E | {\bf{u}}_{ji}^E {\bf{w}}_{ki}^E |^2 \left( | \widetilde {\bf{w}}_{ki}^{A \, \dagger}  {\bf{h}}_{ji}^{A} |^2 + \sigma_{w,i}^2 ||{\bf{h}}_{ji}^{A}||^2  \right ) \right ) \\
\nonumber&& \hspace{5cm} - \log \left ( 1 +  \sum_{i=1}^{N_R} \sum_{k \in \mathcal{M}_i \setminus j} \lambda_{ji}^E | {\bf{u}}_{ji}^E {\bf{w}}_{ki}^E |^2 \left( | \widetilde {\bf{w}}_{ki}^{A \, \dagger}  {\bf{h}}_{ji}^{A} |^2 + \sigma_{w,i}^2 ||{\bf{h}}_{ji}^{A}||^2  \right ) \right ),
\end{eqnarray}
where $\widetilde {\bf{W}}^A = [ \widetilde {\bf{W}}^{A \, T}_1, \dots, \widetilde {\bf{W}}^{A \, T}_{N_R}]^T$ and ${\pmb{\sigma}}_w^2 = [{{\sigma}}_{w,1}^2, \dots, {{\sigma}}_{w,N_R}^2 ]$. 
\subsubsection{Problem Formulation}
As discussed in Section \ref{PFC:HS}, the azimuth precoding $\widetilde {\bf{W}}^A({\bf{H}})$ and the compression noise variance ${\pmb{\sigma}}_w^2 ({\bf{H}})$ can be adapted to the current channel realization at each coherence block. Accordingly, the optimization problem of interest can be formulated as

\begin{subequations} \label{OP_CBP}
\begin{eqnarray} 
\underset { \widetilde {\bf{W}}^A({\bf{H}}), {\bf{W}}^E, \{R_j \}, {\pmb{\sigma}}_{w}^2({\bf{H}}) }{\textrm{maximize}} && \sum_{j \in \mathcal{N}_M} R_j \label{OF_OP;CBP} \\
\textrm{s.t.} \hspace{0.8cm} && R_j \le E [ \bar R_j ({\bf{H}}, \widetilde {\bf{W}}^A({\bf{H}}), {\bf{W}}^E, {\pmb{\sigma}}_{w}^2({\bf{H}}) ) ] , \hspace{2cm} \forall j \in \mathcal{N}_M, \label{RC_OP;CBP} \\
&& C_{w,i} (\widetilde {\bf{W}}_i^A({\bf{H}}), {{\sigma}}_{w,i}^2({\bf{H}})) \le \bar C_i - \sum_{j \in \mathcal{M}_i} R_j, \hspace{1.5cm} \forall i \in \mathcal{N}_R, \label{FC_OP;CBP} \\
&& P_i (\widetilde {\bf{W}}_i^A({\bf{H}}), {\bf{W}}_i^E, {{\sigma}}_{w,i}^2({\bf{H}})) \le \bar P_i, \hspace{2.6cm} \forall i \in \mathcal{N}_R, \label{PC_OP;CBP}
\end{eqnarray}
\end{subequations}
where the constraints apply to every channel realization ${\bf{H}}$. 
\subsubsection{Optimization Algorithm}
Similar to Section \ref{PFC:HS}, the non-convex functions $\bar R_j ({\bf{H}}, \widetilde {\bf{W}}^A({\bf{H}}), {\bf{W}}^E,$  ${\pmb{\sigma}}_{w}^2({\bf{H}}))$ and $C_{w,i} (\widetilde {\bf{W}}_i^A({\bf{H}}), {{\sigma}}_{w,i}^2({\bf{H}}))$ can be seen to be DC functions of the covariance matrices $\widetilde {\bf{V}}^A_{ji}({\bf{H}}) = \widetilde {\bf{w}}_{ji}^A({\bf{H}}) \widetilde{\bf{w}}_{ji}^{A \, \dagger}({\bf{H}})$ and ${\bf{V}}^E_{ji} = {\bf{w}}^E_{ji}{\bf{w}}_{ji}^{E \, \dagger}$ for all $j \in \mathcal{N}_M$ and $i \in \mathcal{N}_R$. Moreover, the optimization problem can be divided into long-term and short-term optimizations, that can be tackled via the SSUM and DC methods, respectively, as summarized in Algorithm \ref{Algorithm_CBP} and detailed in Appendix \ref{Apx;Opt_CBP}. Moreover, as in Algorithm \ref{Algorithm_HS}, it is required to solve one convex problem, which has polynomial complexity \cite{BoydBook}, at each inner iteration.
\begin{figure}[t]
\centering
\includegraphics[width=8cm]{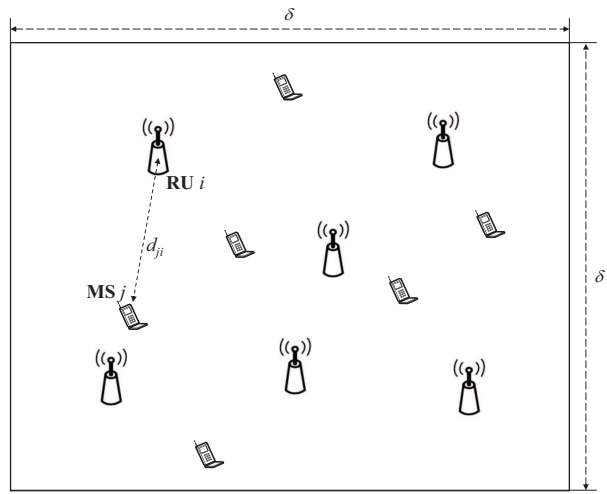}
\caption{Simulation environment for the numerical results.}
\label{fig:Simenv}
\end{figure}
\linespread{1.5}
\begin{algorithm} [t!]
\begin{algorithmic}
\caption{DC Algorithm for Optimization of $\widetilde {\bf{V}}^A({\bf{H}})$ and ${\pmb{\sigma}}_{w}^{2}({\bf{H}})$} \label{Algorithm_DC;CBP}
\State {\textbf{Input:}} Channel ${\bf{H}}$ and elevation precoding ${\bf{V}}^E$.
\State {\textbf{Output:}} $\widetilde {\bf{V}}^A({\bf{H}})$ and ${\pmb{\sigma}}_w^2({\bf{H}})$
\State {\textbf{Initialization}}: Initialize $\widetilde {\bf{V}}^{A \, (0)}({\bf{H}}) \succeq 0$ and ${\pmb{\sigma}}_{w}^{2 \, (0)}({\bf{H}}) \in {\mathbb{R}}^+$, and set $l=0$. 
\State  {\textbf{Repeat}}  
\State \indent $l \gets l+1$ 
\State \indent Update $\widetilde {\bf{V}}^{A \, (l)}({\bf{H}})$ and ${\pmb{\sigma}}_{w}^{2 \, (l)}({\bf{H}})$ by solving problem (\ref{OP_wMM_CBP}).
\State {\textbf{Until}} a convergence criterion is satisfied.
\State Set $\widetilde {\bf{V}}^{A}({\bf{H}}) \gets \widetilde {\bf{V}}^{A \, (l)}({\bf{H}})$ and ${\pmb{\sigma}}_{w}^{2}({\bf{H}}) \gets {\pmb{\sigma}}_{w}^{2 \,\, (l)}({\bf{H}})$.
\end{algorithmic}
\end{algorithm}
\linespread{2}
\section{Numerical Results}
In this section, we compare the performance of the strategies with layered precoding, namely layered CAP and CBP schemes, and the conventional strategies, namely CAP and CBP schemes, for FD-MIMO systems. To this end, we consider a set-up simulation environment where the RUs and MSs are randomly located in a square area with side $\delta = 500$ m as in Fig. \ref{fig:Simenv}. In the path loss formula (\ref{PL_coef}), we set the reference distance to $d_0=50$ m and the path loss exponent to $\eta = 3$ with $d_{ji}$ being the Euclidean distance between the $i$-th RU and the $j$-th MS. The channels are assumed to have the Kronecker model in (\ref{CH_Model}). Throughout, we assume that the every RU is subject to the same fronthaul capacity $\bar C$ and has the same power constraint $\bar P$, namely $\bar C_i = \bar C$ and $\bar P_i = \bar  P$ for $i \in \mathcal{N}_R$. Throughout, we consider CBP strategies in which each RU serves all MSs, i.e., $N_C=N_M$.

\begin{figure}[t]
\centering
\vspace{-0.5cm}
\includegraphics[width=14cm]{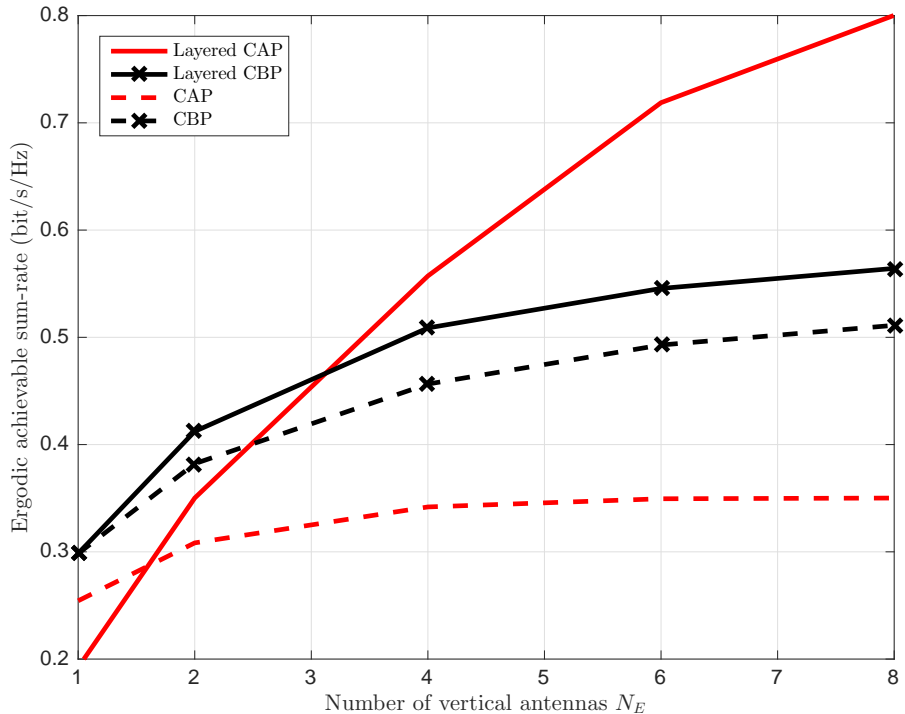}
\vspace{-0.5cm}
\caption{Ergodic achievable sum-rate vs. the number of vertical antennas $N_E$ ($N_R = N_M = 2$, $N_{A,i} = 2$, $C=1$ bit/s/Hz, $P = 0$ dB, and $T = 20$).}
\label{fig:fig6}
\end{figure}

\begin{figure}[t]
\centering
\vspace{-0.5cm}
\includegraphics[width=14cm]{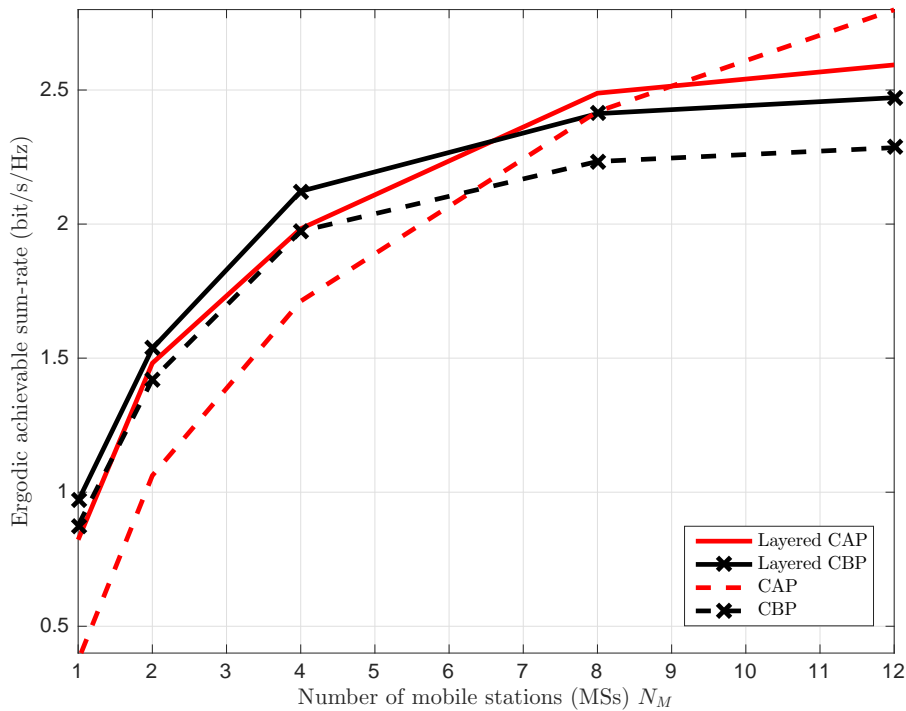}
\vspace{-0.5cm}
\caption{Ergodic achievable sum-rate vs. the number of MSs $N_M$ ($N_R = 2$, $N_{A,i} = 2$, $N_{E,i} = 4$, $C=3$ bit/s/Hz, $ P = 5$ dB, and $T = 20$).}
\label{fig:vsNM}
\end{figure}

Fig. \ref{fig:fig6} shows the ergodic achievable sum-rate as function of the number of vertical antennas $N_E$, where the number of RUs and MSs is $N_R=N_M=2$, the number of horizontal antennas is $N_{A,i} = 2$ for all $i \in {\mathcal{N}_R}$, the fronthaul capacity is $\bar C = 1$ bit/s/Hz, the transmit power is $\bar P = 0$ dB and the coherence time is $T=20$. We observe that the layered precoding schemes provide increasingly large gains as $N_E$ grows larger. This is because, in the conventional strategies, the fronthaul overhead for the transfer of elevation precoding information increases with the number of vertical antennas. This gain is less pronounced here for layered CBP strategies, whose achievable rate is limited here by the relatively small coherence interval, as further discussed below (see also Sec. \ref{CBP}). Moreover, it is observed that, for $N_E=1$, the conventional and the layered precoding strategies with CBP method have the same performance, while this is not the case for the CAP strategies. In fact, the the conventional CAP strategy outperforms the layered CAP strategy for small values of $N_E$. This is caused by the fact that, with the layered CAP strategy, the azimuth precoded signals for the MSs are separately compressed, hence entailing an inefficient use of the fronthaul when $N_E$ is large enough.

Fig. \ref{fig:vsNM} shows the effect of the number of MSs $N_M$ on the ergodic achievable sum-rate with $N_R=2$, $N_{A,i}=2$ and $N_{E,i}=4$ for all $i \in \mathcal{N}_R$, $C=3$ bit/s/Hz, $P=5$ dB, and $T=20$. The CBP methods show the known poor performance as the number of MSs increases, due to the need for the transmission of the messages of all MSs on all fronthaul links \cite{Kang14arXiv}. Moreover, in keeping with the discussion above, we observe that the conventional CAP method is to be preferred in the regime of large number of MSs. This is due to the separate compression of the azimuth precoded signals of layered CAP, which entails a fronthaul overhead proportional to the number of MSs. 

\begin{figure}[t]
\centering
\vspace{-0.5cm}
\includegraphics[width=14cm]{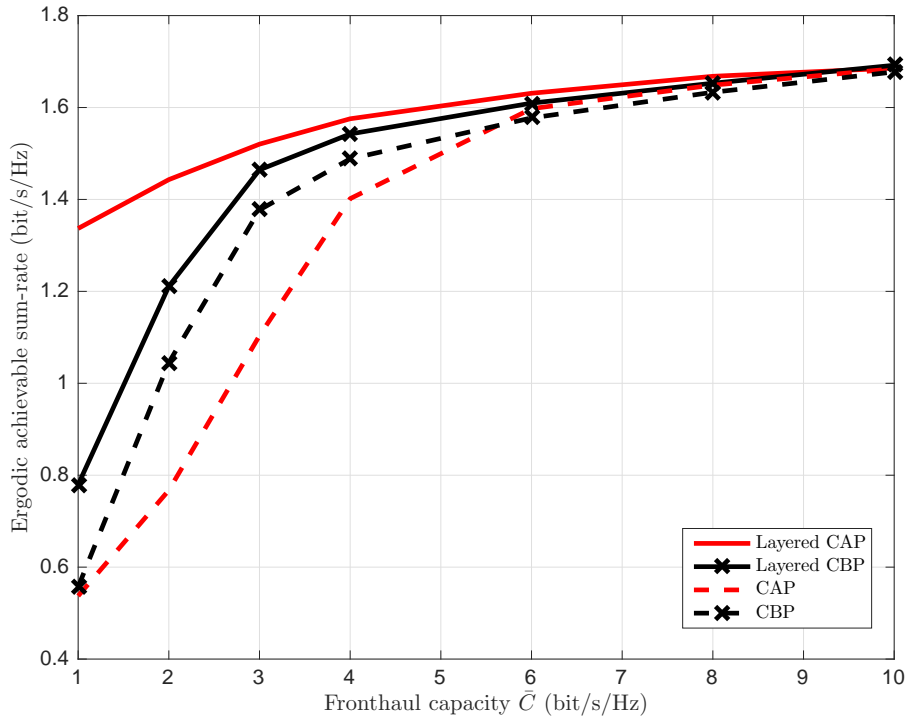}
\vspace{-0.5cm}
\caption{Ergodic achievable sum-rate vs. the fronthaul capacity $\bar C$ ($N_R =2$, $N_M = 2$, $N_{A,i} = 2$, $N_{E,i} = 4$, $P = 5$ dB, and $T = 10$).}
\label{fig:fig7}
\end{figure}

\begin{figure}[h!]
\centering
\vspace{-0.5cm}
\includegraphics[width=14cm]{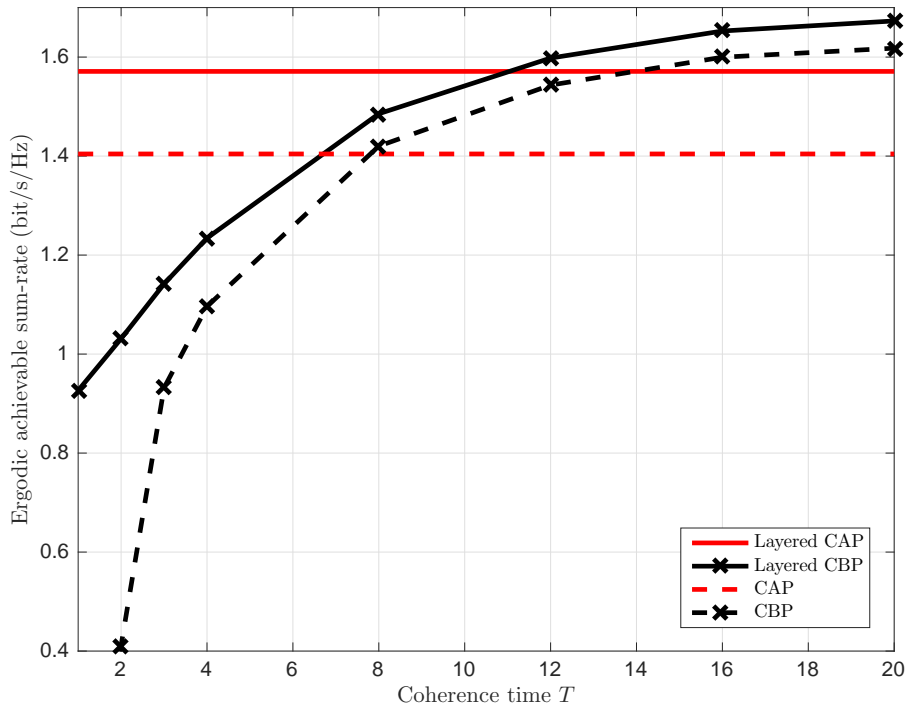}
\vspace{-0.5cm}
\caption{Ergodic achievable sum-rate vs. the coherence time $T$ ($N_R =N_M = 2$, $N_{A,i} = 2$, $N_{E,i} = 4$, $C=4$ bit/s/Hz, and $P = 5$ dB).}
\label{fig:fig8}
\end{figure}

In Fig. \ref{fig:fig7}, the ergodic achievable sum-rate is plotted versus the fronthaul capacity $\bar C$ for $N_R=N_M=2$, $N_{A,i} = 2$ and $N_{E,i}=4$ for all $i \in {\mathcal{N}_R}$, $\bar P = 5$ dB, and $T=10$. We first remark that the performance gain of the layered strategies is observed at low-to-moderate fronthaul capacities, while, for large fronthaul capacities, the performance of the conventional strategies approach that of the layered strategies. As a general rule, the conventional CAP strategy is uniformly better than conventional CBP as long as the fronthaul capacity is sufficiently large, due to the enhanced interference mitigation capabilities of CAP \cite{Kang14arXiv}. Instead, the layered CAP strategy is advantageous here across all values of fronthaul capacity. 

The effect of the coherence time $T$ is investigated in Fig. \ref{fig:fig8}, with $N_R=N_M=2$, $N_{A,i} = 2$ and $N_{E,i}=4$ for all $i \in {\mathcal{N}_R}$, $\bar C = 4$ bit/s/Hz, and $\bar P = 5$ dB. The CBP schemes benefit from a larger coherence time $T$, since the fronthaul overhead required to transmit precoding information gets amortized over a larger period. In contrast, such overhead in layered CAP and CAP schemes scales proportionally to the coherence time $T$ and hence the layered CAP and CAP schemes are not affected by the coherence time. 
\section{Concluding Remarks}
In this paper, we have studied the design of downlink Cloud Radio Access Network (C-RAN) systems in which the Radio Units (RUs) are equipped with Full Dimensional (FD)-MIMO arrays. We proposed to leverage the special low-rank structure of FD-MIMO channel, which exhibits different rates of variability in the elevation and azimuth components, by means of a novel layered precoding strategy coupled with an adaptive fronthaul compression scheme. Specifically, in the layered strategy, a single precoding matrix is optimized for the elevation channel across all coherence times based on long-term Channel State Information (CSI), while azimuth precoding matrices are optimized across independent coherence interval by adapting to instantaneous CSI. This proposed layered approach has the unique advantage in a C-RAN of potentially reducing the fronthaul overhead, due to the opportunity to amortize the overhead related to the elevation channel component across multiple coherence times. Via numerical results, it is shown that the layered strategies significantly outperform standard non-layered schemes, especially in the regime of low fronthaul capacity and large number of vertical antennas. 

We have also considered two different functional splits for both layered and non-layered precoding, namely the conventional C-RAN implementation, also known as Compress-After-Precoding (CAP) scheme, and an alternative split, referred to as Compress-Before-Precoding (CBP), whereby channel coding and precoding are performed at the RUs. Layered precoding is seen to work better under a CAP implementation when the coherence interval is not too large and the number of vertical antennas is sufficiently large; whereas the CBP approach benefits from a longer coherence interval due to its capability to amortize the fronthaul overhead for transfer of azimuth precoding information. Interesting open issues include the investigation of a scenario with multiple interfering clusters of RUs controlled by distinct Central Units (CUs) (see \cite{Park14WCL}), and the analysis of the performance in the presence of more general FD-MIMO channel models (see, e.g., \cite{Xu14CTW}).
\appendices
\section{Optimization Algorithm for the Layered CAP Strategy} 
\label{Apx;Opt_CAP}
In this Appendix, we detail the derivation of Algorithm \ref{Algorithm_HS} for the optimization of the layered CAP strategy. We first discuss the optimization problem for the short-term variables, namely the covariance matrix ${\bf{V}}^{A }({\bf{H}})$ for azimuth precoding and the quantization noise variance ${\pmb{\sigma}}_{x}^{2} ({\bf{H}})$, which are adapted to the channel realization ${\bf{H}}$, for given the elevation covariance matrix ${\bf{V}}^E$. We then consider the optimization of the long-term variable, namely the covariance matrix ${\bf{V}}^{E}$ for elevation precoding, with the given covariance matrices ${\bf{V}}^{A}({\bf{H}})$ for azimuth precoding and quantization noise vectors ${\pmb{\sigma}}_{x}^{2} ({\bf{H}})$. 

After obtaining the elevation covariance matrix ${{\bf{V}}^{E}}^{{\pmb{*}}}$, using the approach in Algorithm \ref{Algorithm_HS}, the precoding matrix ${{\bf{W}}^{E}}^{{\pmb{*}}}$ for the elevation channel is calculated via the principal eigenvector approximation \cite{BoydSemiProg} of the obtained solution ${{\bf{V}}^{E}}^{{\pmb{*}}}$ as ${{\bf{w}}^E_{ji}}^{{\pmb{*}}}  = {\mathbf{\nu}}_{\textrm{max}} ({{\bf{V}}_{ji}^{E}}^{\pmb{*}})$ for all $j \in \mathcal{N}_M$ and $i \in \mathcal{N}_R$. In a similar fashion, the algorithm obtains the precoding matrix ${{\bf{W}}^{A}}^{{\pmb{*}}} ({\bf{H}})$ for the azimuth channel via the standard rank-reduction approach \cite{BoydSemiProg} from the obtained solution ${{\bf{V}}^{A} ({\bf{H}})}^{{\pmb{*}}}$ as ${{\bf{w}}^A_{ji}}^{{\pmb{*}}} ({\bf{H}}) = \beta_{ji} {\mathbf{\nu}}_{\textrm{max}} ( {{\bf{V}}_{ji}^{A}({\bf{H}})}^{{\pmb{*}}} )$ with the normalization factors $\beta_{ji}$ selected to satisfy the power constraint with equality, namely $P_i ({{\bf{W}}_i^A}^{{\pmb{*}}} ({\bf{H}}), {{\bf{W}}_i^E}^{{\pmb{*}}}, {{\pmb{\sigma}}_{x,i}^2}^{{\pmb{*}}} ({\bf{H}})) = \bar P_i$. 
\subsection{Optimization over ${\bf{V}}^{A}({\bf{H}})$ and ${\pmb{\sigma}}_{x}^{2} ({\bf{H}})$ with given ${\bf{V}}^E$} \label{OP_H_JOP_HS} 
Here, we tackle the problem (\ref{OP_HS}) based on the DC algorithm \cite{MMBook} given the elevation precoding covariance matrix ${\bf{V}}^E$ over the azimuth covariance matrix ${\bf{V}}^{A}({\bf{H}})$ and the quantization noise variance ${\pmb{\sigma}}_{x}^{2} ({\bf{H}})$. To this end, the objective function $R_j ({\bf{H}}, {\bf{W}}^A({\bf{H}}), {\bf{W}}^E, {\pmb{\sigma}}_{x}^2({\bf{H}}))$ is approximated by a locally tight lower bound $\widetilde R_j ({\bf{H}}, {\bf{V}}^{A}({\bf{H}}),$ ${\pmb{\sigma}}_{x}^{2}({\bf{H}}) | {\bf{V}}^{A \, (l-1)}({\bf{H}}), {\pmb{\sigma}}_{x}^{2 \, (l-1)}({\bf{H}}), {\bf{V}}^{E})$ around solutions ${\bf{V}}^{A \, (l-1)}({\bf{H}})$ and ${\pmb{\sigma}}_{x}^{2 \, (l-1)} ({\bf{H}})$ obtained at $(l-1)$-th inner iteration with
\begin{eqnarray} \label{AR_wMM_HS}
 && \hspace{-1.3cm} \widetilde R_j ({\bf{H}}, {\bf{V}}^{A}({\bf{H}}), {\pmb{\sigma}}_{x}^{2}({\bf{H}}) | {\bf{V}}^{A \, (l-1)}({\bf{H}}), {\pmb{\sigma}}_{x}^{2 \, (l-1)}({\bf{H}}), {\bf{V}}^{E }) = \log \left( 1 + \sum_{k=1}^{N_M} \sum_{i=1}^{N_R} \rho_{ji} ({\bf{V}}_{ki}^A({\bf{H}}), {\bf{V}}_{ki}^{E}, \sigma_{x,ki}^2({\bf{H}})) \right ) \\
\nonumber&& \hspace{-0.5cm} - f \hspace{-0.1cm} \left( \hspace{-0.1cm} 1 \hspace{-0.1cm} + \hspace{-0.3cm} \sum_{k=1,k \neq j}^{N_M} \sum_{i=1}^{N_R} \hspace{-0.1cm} \rho_{ji} ({\bf{V}}_{ki}^{\hspace{-0.05cm}A \, (l-1)}\hspace{-0.05cm}({\bf{H}}), {\bf{V}}_{ki}^{\hspace{-0.05cm} E}, {\pmb{\sigma}}_{x,ki}^{2 \, (l-1)}\hspace{-0.05cm}({\bf{H}})), 1 \hspace{-0.1cm} + \hspace{-0.3cm}\sum_{k=1,k \neq j}^{N_M} \sum_{i=1}^{N_R} \rho_{ji} ({\bf{V}}_{ki}^A({\bf{H}}), {\bf{V}}_{ki}^{E}, \sigma_{x,ki}^2({\bf{H}})) \hspace{-0.1cm} \right)
\end{eqnarray}
where $\rho_{ji} ({\bf{V}}_{ki}^A, {\bf{V}}_{ki}^E, \sigma_{x,ki}^2) = \lambda_{ji}^E {\bf{u}}_{ji}^E {\bf{V}}_{ki}^E {\bf{u}}^\dagger_{ji} \left( {\bf{h}}_{ji}^{A} {\bf{V}}_{ki}^{A} {\bf{h}}_{ji}^{A \, \dagger} + \sigma_{x,ki}^2 ||{\bf{h}}_{ji}^{A}||^2  \right )$ and the linearized function $f(a,b)$ is obtained from the first-order Taylor expansion of the log function as $f(a,b) = \log (a) + {(b-a)}/{a}$. Since the fronthaul constraint (\ref{FC_OP;HS}) is a DC constraint, the left-hand side of the constraint (\ref{FC_OP;HS}) is approximated by applying successive locally tight convex lower bounds as
\begin{eqnarray}
&& \widetilde C_{x,i} ({\bf{V}}_i^A({\bf{H}}), {\pmb{\sigma}}_{x,i}^2({\bf{H}}) | {\bf{V}}_i^{A \, (l-1)}({\bf{H}}), {\pmb{\sigma}}_{x,i}^{2 \, (l-1)}({\bf{H}})) \triangleq \\
\nonumber && \hspace{4cm} \sum_{j=1}^{N_M} \left \{ f \left ({\textrm{tr}}({\bf{V}}_{ji}^{A \, (l-1)}({\bf{H}})) + \sigma_{x,ji}^{2 \, (l-1)}({\bf{H}}), {\textrm{tr}}({\bf{V}}_{ji}^A({\bf{H}})) + \sigma_{x,ji}^{2}({\bf{H}}) \right) - \log \sigma_{x,ji}^{2} \right \}.
\end{eqnarray}
At $l$-th inner loop, the following convex optimization problem, for given ${\bf{V}}^{A \, (l-1)}({\bf{H}})$, ${\pmb{\sigma}}_{x}^{2 \, (l-1)} ({\bf{H}})$, and ${\bf{V}}^{E}$, is solved for obtaining new iterates ${\bf{V}}^{A \, (l)}({\bf{H}})$ and ${\pmb{\sigma}}_{x}^{2 \, (l)} ({\bf{H}})$ as 
\begin{subequations} \label{OP_wMM_HS}
\begin{eqnarray} 
\hspace{-0.7cm} {\bf{V}}^{A \, (l)}({\bf{H}}), {\pmb{\sigma}}_{x}^{2 \, (l)} ({\bf{H}}) \gets \hspace{-0.2cm} \underset { {\bf{V}}^A({\bf{H}}), {\pmb{\sigma}}_{x}^2({\bf{H}})}{\textrm{arg max}} && \hspace{-0.6cm} \sum_{j \in \mathcal{N}_M} \hspace{-0.1cm} \widetilde R_j ({\bf{H}}, {\bf{V}}^{A}({\bf{H}}), {\pmb{\sigma}}_{x}^{2}({\bf{H}}) | {\bf{V}}^{A \, (l-1)}({\bf{H}}), \hspace{-0.05cm} {\pmb{\sigma}}_{x}^{2 \, (l-1)}({\bf{H}}), \hspace{-0.05cm} {\bf{V}}^{E}) \label{OF_OP_wMM;HS} \\
\textrm{s.t.} \hspace{0.7cm} && \hspace{-0.6cm} \widetilde C_{x,i} ({\bf{V}}_i^A({\bf{H}}), {\pmb{\sigma}}_{x,i}^2({\bf{H}}) | {\bf{V}}_i^{A \, (l-1)}({\bf{H}}), {\pmb{\sigma}}_{x,i}^{2 \, (l-1)}({\bf{H}})) \le \bar C_i, \label{FC_OP_wMM;HS} \\
&& \hspace{-0.6cm} P_i ({\bf{V}}_i^A({\bf{H}}), {\bf{V}}_i^{E}, {\pmb{\sigma}}_{x,i}^2({\bf{H}})) \le \bar P_i, \hspace{1cm} \forall i \in \mathcal{N}_R. \label{PC_OP_wMM;HS}
\end{eqnarray}
\end{subequations}
The DC method obtains the solutions ${\bf{V}}^{A}({\bf{H}})$ and ${\pmb{\sigma}}_{x}^{2} ({\bf{H}})$ by solving the problem (\ref{OP_wMM_HS}) iteratively over $l$ until a convergence criterion is satisfied and the resulting algorithm is summarized in Algorithm \ref{Algorithm_DC}.
\subsection{Optimization over ${\bf{V}}^{E}$} \label{OP_JOP_HS}
In this part, the covariance matrix ${\bf{V}}^{E}$ for elevation precoding is designed for given azimuth precoding covariance matrices ${\bf{V}}^{A \, (m)} = {\bf{V}}^{A \, (m)}({\bf{H}}^{(m)})$ and quantization noise vectors ${\pmb{\sigma}}_{x}^{2 \, (m)} = {\pmb{\sigma}}_{x}^{2 \, (m)} ({\bf{H}}^{(m)})$ for all $m=1,\dots, n$. Since the elevation covariance matrix ${\bf{V}}^{E \, (n)}$ is not adapted to the channel realization ${\bf{H}}$ and the objective function (\ref{OP_HS}) is non-convex with respect to ${\bf{V}}^{E \, (n)}$, in this optimization, we use the SSUM algorithm \cite{SSUM_paper}. To this end, at each step, a stochastic lower bound of the objective function is maximized around the current iterate. Following the SSUM method, at $n$-th outer loop, the objective function with given ${\bf{V}}^{A \, (m)}$ and ${\pmb{\sigma}}_{x}^{2 \, (m)}$, for all $m=1,\dots, n$, is reformulated as the empirical average
\begin{equation}
\frac{1}{n} \sum_{m=1}^{n} \widetilde R_j ({\bf{H}}^{(m)}, {\bf{V}}^E | {\bf{V}}^{E \, (m-1)}, {\bf{V}}^{A \, (m)}, {\pmb{\sigma}}_{x}^{2 \, (m)}),
\end{equation}
where $\widetilde R_j ({\bf{H}}^{(m)}, {\bf{V}}^E | {\bf{V}}^{E \, (m-1)}, {\bf{V}}^{A}, {\pmb{\sigma}}_{x}^{2})$ is a locally tight convex lower bound around the previous iterate ${\bf{V}}^{E \, (m-1)}$, when the channel realization is ${\bf{H}}^{(m)}$, and is calculated as
\begin{eqnarray} \label{AR_wSSUM_HS}
&& \hspace{-0.5cm} \widetilde R_j ({\bf{H}}^{(m)}, {\bf{V}}^E | {\bf{V}}^{E \, (m-1)}, {\bf{V}}^{A \, (m)}, {\pmb{\sigma}}_{x}^{2 \, (m)}) = \log \left( 1 + \sum_{k=1}^{N_M} \sum_{i=1}^{N_R} \rho_{ji} ({\bf{H}}^{(m)}, {\bf{V}}_{ki}^{A \, (m)}, {\bf{V}}_{ki}^E, \sigma_{x,i}^{2 \, (m)}) \right ) \\
\nonumber&& \hspace{-0.5cm} - f \left( 1 + \sum_{k=1,k \neq j}^{N_M} \sum_{i=1}^{N_R} \rho_{ki} ({\bf{H}}^{(m)}, {\bf{V}}_{ki}^{A \, (m)}, {\bf{V}}_{ki}^{E \, (m-1)}, {\pmb{\sigma}}_{x,i}^{2 \, (m)}), 1 + \sum_{k=1,k \neq j}^{N_M} \sum_{i=1}^{N_R} \rho_{ki} ({\bf{H}}^{(m)}, {\bf{V}}_{ki}^{A \, (m)}, {\bf{V}}_{ki}^E, \sigma_{x,i}^{2 \, (m)}) \right),
\end{eqnarray}
with $\rho_{ji} ({\bf{H}}^{(m)}, {\bf{V}}_{ki}^A, {\bf{V}}_{ki}^E, \sigma_{x,i}^2) = \lambda_{ji}^{E \, (m)} {\bf{u}}_{ji}^{E \, (m)} {\bf{V}}_{ki}^E {\bf{u}}^{(m) \, \dagger}_{ji} ( {\bf{h}}_{ji}^{A \, (m)} {\bf{V}}_{ki}^{A} {\bf{h}}_{ji}^{A  \, (m) \, \dagger} + \sigma_{x,i}^2 ||{\bf{h}}_{ji}^{A \, (m)}||^2   )$. The $n$-th iterate ${\bf{V}}^{E \, (n)}$ is obtained by solving the following convex optimization problem
\begin{subequations} \label{OP_wSSUM_HS}
\begin{eqnarray} 
{\bf{V}}^{E \, (n)} \gets \underset {{\bf{V}}^E}{\textrm{arg max}} && \frac{1}{n} \sum_{m=1}^{n} \sum_{j \in \mathcal{N}_M} \widetilde R_j ({\bf{H}}^{(m)}, {\bf{V}}^E | {\bf{V}}^{E \, (m-1)}, {\bf{V}}^{A \, (m)}, {\pmb{\sigma}}_{x}^{2 \, (m)}) \label{OF_OP_wSSUM;HS} \\
\textrm{s.t.} && C_{x,i} ({\bf{V}}_i^{A \, (n)}, {\pmb{\sigma}}_{x,i}^{2 \, (n)}) \le \bar C_i, \hspace{1cm} \forall i \in \mathcal{N}_R, \label{FC_OP_wSSUM;HS} \\
&& P_i ({\bf{V}}_i^{A \, (n)}, {\bf{V}}_i^E, {\pmb{\sigma}}_{x,i}^{2 \, (n)}) \le \bar P_i, \hspace{1cm} \forall i \in \mathcal{N}_R. \label{PC_OP_wSSUM;HS}
\end{eqnarray}
\end{subequations}
As in Section \ref{OP_H_JOP_HS}, the outer loop in Algorithm \ref{Algorithm_HS} is repeated until the convergence is achieved.
\section{Optimization Algorithm for Layered CBP Strategy} \label{Apx;Opt_CBP}
In this Appendix, the precoding matrices ${{\bf{W}}^{E}}^{{\pmb{*}}}$ and $\widetilde {\bf{W}}^{A \, {{\pmb{*}}}}$, MSs' rates $\{ R_j \}$ and quantization noise vector ${{\pmb{\sigma}}_w^2}^{{\pmb{*}}}$ are jointly optimized for the CBP-based strategy.
The optimization of short-term variables, namely the covariance matrix $\widetilde {\bf{V}}^{A}({\bf{H}})$ for azimuth precoding and the quantization noise variance ${\pmb{\sigma}}_{w}^{2} ({\bf{H}})$, which are adapted to the channel realization ${\bf{H}}$ for given the elevation covariance matrix ${\bf{V}}^E$, is described first. Then, the optimization over the long-term variables, namely the covariance matrix ${\bf{V}}^{E}$ for elevation precoding and the user rates $\{R_j\}$, is discussed given covariance matrices ${\bf{V}}^{A \, (m)}({\bf{H}})$ for azimuth precoding and quantization noise vectors ${\pmb{\sigma}}_{w}^{2 \, (m)} ({\bf{H}})$, for all $m=1,\dots, n$, as detailed in Appendix \ref{OP_JOP_CBP}. 

As in Appendix \ref{Apx;Opt_CAP}, the elevation precoding matrix ${{\bf{W}}^{E}}^{\pmb{*}}$ and the azimuth precoding matrix $\widetilde {\bf{W}}^{A \, {\pmb{*}}}$ are calculated via the standard rank-reduction approach \cite{BoydSemiProg} with the obtained solutions ${{\bf{V}}^{E}}^{\pmb{*}}$ and $\widetilde {\bf{V}}^{A \, {\pmb{*}}}$, respectively, as detailed in Algorithm \ref{Algorithm_CBP}.
\subsection{Optimization over $\widetilde {\bf{V}}^{A}({\bf{H}})$ and ${\pmb{\sigma}}_{w}^{2} ({\bf{H}})$ with given ${\bf{V}}^E$} \label{OP_H_JOP_CBP} 
Here, we aim at maximizing the objective function (\ref{OF_OP;CBP}) over the azimuth precoding covariance matrix $\widetilde {\bf{V}}^{A}({\bf{H}})$ and the quantization noise variance ${\pmb{\sigma}}_{w}^{2} ({\bf{H}})$ given the elevation precoding covariance matrix ${\bf{V}}^E$ using the DC method \cite{MMBook}. At the $l$-th iteration of the DC method, the non-convex functions $\bar R_j ({\bf{H}}, \widetilde {\bf{V}}^A({\bf{H}}), {\bf{V}}^E, {\pmb{\sigma}}_{w}^2({\bf{H}}))$ and $C_{w,i} (\widetilde {\bf{V}}_i^A({\bf{H}}), {{\sigma}}_{w,i}^2({\bf{H}}))$ are respectively substituted with a locally tight lower bound $\widetilde R_j ({\bf{H}}, \widetilde {\bf{V}}^A({\bf{H}}), {\pmb{\sigma}}_{w}^2({\bf{H}}) |$ $\widetilde {\bf{V}}^{A \, (l-1)}({\bf{H}}), {\pmb{\sigma}}_{w}^{2 \, (l-1)}({\bf{H}}), {\bf{V}}^E)$ and a tight upper bound $\widetilde C_{w,i} (\widetilde {\bf{V}}_i^A({\bf{H}}),$ ${{\sigma}}_{w,i}^2({\bf{H}}) |\widetilde {\bf{V}}_i^{A \, (l-1)}({\bf{H}}), {{\sigma}}_{w,i}^{2 \, (l-1)}({\bf{H}}))$, obtained as in Appendix \ref{Apx;Opt_CAP}. The bounds are given by 
\begin{eqnarray} 
\nonumber && \hspace{-0.8cm} \widetilde R_j ({\bf{H}}, \widetilde {\bf{V}}^A({\bf{H}}), {\pmb{\sigma}}_{w}^2({\bf{H}}) | \widetilde {\bf{V}}^{A \, (l-1)}({\bf{H}}), {\pmb{\sigma}}_{w}^{2 \, (l-1)}({\bf{H}}), {\bf{V}}^E) = \log \left( 1 + \sum_{i=1}^{N_R} \sum_{k \in \mathcal{M}_i} \rho_{ji} ( \widetilde {\bf{V}}_{ki}^A({\bf{H}}), {\bf{V}}_{ki}^{E}, \sigma_{w,i}^2({\bf{H}})) \right ) \\
&& \hspace{-0.1cm} - f \hspace{-0.1cm} \left( \hspace{-0.1cm} 1 \hspace{-0.1cm} + \sum_{i=1}^{N_R} \sum_{k \in \mathcal{M}_i \setminus j} \hspace{-0.1cm} \rho_{ki} (\widetilde {\bf{V}}_{ki}^{\hspace{-0.05cm}A \, (l-1)}\hspace{-0.05cm}({\bf{H}}), {\bf{V}}_{ki}^{\hspace{-0.05cm} E}, {{\sigma}}_{w,i}^{2 \, (l-1)}\hspace{-0.05cm}({\bf{H}})), 1 \hspace{-0.1cm} + \sum_{i=1}^{N_R} \sum_{k \in \mathcal{M}_i \setminus j} \rho_{ki} (\widetilde {\bf{V}}_{ki}^A({\bf{H}}), {\bf{V}}_{ki}^{E}, \sigma_{w,i}^2({\bf{H}})) \hspace{-0.1cm} \right),
\end{eqnarray}
and 
\begin{eqnarray}
&& \hspace{-2cm} \widetilde C_{w,i} (\widetilde {\bf{V}}_i^A({\bf{H}}), {{\sigma}}_{w,i}^2({\bf{H}}) | \widetilde {\bf{V}}_i^{A \, (l-1)}({\bf{H}}), {{\sigma}}_{w,i}^{2 \, (l-1)}({\bf{H}})) \triangleq \\
\nonumber && \hspace{1cm} \frac{1}{T} \left \{ f \left ( \widetilde {\bf{V}}_i^{A \, (l-1)}({\bf{H}}) + {{\sigma}}_{w,i}^{2 \, (l-1)}({\bf{H}}) {\bf{I}}, \widetilde {\bf{V}}_i^{A}({\bf{H}}) + {{\sigma}}_{w,i}^2({\bf{H}}) {\bf{I}} \right) - N_{A,i} \log \left ( \sigma_{w,i}^2 \right) \right \},
\end{eqnarray}
where  $\rho_{ji} (\widetilde {\bf{V}}_{ki}^A, {\bf{V}}_{ki}^E, \sigma_{w,i}^2) = \lambda_{ji}^E {\bf{u}}_{ji}^E {\bf{V}}_{ki}^E {\bf{u}}^\dagger_{ji} \left( {\bf{h}}_{ji}^{A} \widetilde {\bf{V}}_{ki}^{A} {\bf{h}}_{ji}^{A \, \dagger} + \sigma_{w,i}^2 ||{\bf{h}}_{ji}^{A}||^2  \right )$ and the linearization function $f({\bf{A}},{\bf{B}})$ for the matrices is defined as $f ({\bf{A}},{\bf{B}}) \triangleq \log \det ({\bf{A}}) + \textrm{tr} ({\bf{A}}^{-1} ({\bf{B}} - {\bf{A}}))$.

At $l$-th iteration of DC method, the following convex optimization problem for given $\widetilde {\bf{V}}^{A \, (l-1)}({\bf{H}})$, ${\pmb{\sigma}}_{w}^{2 \, (l-1)} ({\bf{H}})$ and ${\bf{V}}^{E}$ is solved for obtaining new iterates $\widetilde {\bf{V}}^{A \, (l)}({\bf{H}})$ and ${\pmb{\sigma}}_{w}^{2 \, (l)} ({\bf{H}})$:
\begin{subequations} \label{OP_wMM_CBP}
\begin{eqnarray} 
\hspace{-0.5cm} \widetilde {\bf{V}}^{A \, (l)}({\bf{H}}), {\pmb{\sigma}}_{w}^{2 \, (l)} ({\bf{H}}) \gets && \hspace{-0.5cm} \underset { \widetilde {\bf{V}}^A({\bf{H}}), {\pmb{\sigma}}_{w}^2({\bf{H}}), \{R_j\}}{\textrm{arg max}} \sum_{j \in \mathcal{N}_M} \hspace{-0.1cm} R_j \label{OF_OP_wMM;CBP} \\
\textrm{s.t.} && \hspace{0.2cm}  R_j \le \widetilde R_j ({\bf{H}}, \widetilde {\bf{V}}^A({\bf{H}}), {\pmb{\sigma}}_{w}^2({\bf{H}}) | \widetilde {\bf{V}}^{A \, (l-1)}({\bf{H}}), {\pmb{\sigma}}_{w}^{2 \, (l-1)}({\bf{H}}), {\bf{V}}^E), \,\, \forall j \in \mathcal{N}_M, \label{RC_OP_wMM;CBP} \\
&& \hspace{0.2cm} \widetilde C_{w,i} (\widetilde {\bf{V}}_i^A({\bf{H}}), {{\sigma}}_{w,i}^2({\bf{H}}) | \widetilde {\bf{V}}_i^{A \, (l-1)}({\bf{H}}), {{\sigma}}_{w,i}^{2 \, (l-1)}({\bf{H}})) \le \bar C_i - \sum_{j \in \mathcal{M}_i} R_j, \label{FC_OP_wMM;CBP} \\
&& \hspace{0.2cm} P_i (\widetilde {\bf{V}}_i^A({\bf{H}}), {\bf{V}}_i^{E}, {{\sigma}}_{w,i}^2({\bf{H}})) \le \bar P_i, \hspace{1cm} \forall i \in \mathcal{N}_R. \label{PC_OP_wMM;CBP}
\end{eqnarray}
\end{subequations}
Problem (\ref{OP_wMM_CBP}) is solved iteratively over $l$ until convergence and the resulting algorithm is summarized in Algorithm \ref{Algorithm_DC;CBP}.

\subsection{Optimization over ${\bf{V}}^{E}$ and $\{ R_j \}$} \label{OP_JOP_CBP}
We design the covariance matrix ${\bf{V}}^{E}$ for elevation precoding and the user rates $\{R_j\}$ for given azimuth precoding covariance matrices $\widetilde {\bf{V}}^{A \, (m)} = \widetilde{\bf{V}}^{A \, (m)}({\bf{H}}^{(m)})$ and quantization noise vectors ${\pmb{\sigma}}_{w}^{2 \, (m)} = {\pmb{\sigma}}_{w}^{2 \, (m)} ({\bf{H}}^{(m)})$ for all $m=1,\dots, n$. As in Appendix \ref{Apx;Opt_CAP}, this optimization problem can be tackled via the SSUM method. To this end, the function $E [ \bar R_j ({\bf{H}}, \widetilde {\bf{W}}^A({\bf{H}}), {\bf{W}}^E, {\pmb{\sigma}}_{w}^2({\bf{H}}) )]$ in (\ref{RC_OP;CBP}) is approximated with the stochastic upper bound as
\begin{equation}
\frac{1}{n} \sum_{m=1}^{n} \widetilde R_j ({\bf{H}}^{(m)}, {\bf{V}}^E | {\bf{V}}^{E \, (m-1)}, \widetilde {\bf{V}}^{A \, (m)}, {\pmb{\sigma}}_{w}^{2 \, (m)}),
\end{equation}
with
\begin{eqnarray} \label{AR_wSSUM_CBP}
&& \hspace{-0.8cm} \widetilde R_j ({\bf{H}}^{(m)}, {\bf{V}}^E | {\bf{V}}^{E \, (m-1)}, \widetilde {\bf{V}}^{A \, (m)}, {\pmb{\sigma}}_{w}^{2 \, (m)}) = \log \left( 1 + \sum_{i=1}^{N_R} \sum_{k \in \mathcal{M}_i} \rho_{ji} ({\bf{H}}^{(m)}, \widetilde {\bf{V}}_{ki}^{A \, (m)}, {\bf{V}}_{ki}^E, \sigma_{w,i}^{2 \, (m)}) \right ) \\
\nonumber&& \hspace{-0.5cm} - f \left( 1 + \sum_{i=1}^{N_R} \sum_{k \in \mathcal{M}_i \setminus j} \rho_{ki} ({\bf{H}}^{(m)}, \widetilde {\bf{V}}_{ki}^{A \, (m)}, {\bf{V}}_{ki}^{E \, (m-1)}, {{\sigma}}_{w,i}^{2 \, (m)}), 1 + \sum_{i=1}^{N_R} \sum_{k \in \mathcal{M}_i \setminus j} \rho_{ki} ({\bf{H}}^{(m)}, \widetilde {\bf{V}}_{ki}^{A \, (m)}, {\bf{V}}_{ki}^E, \sigma_{w,i}^{2 \, (m)}) \right),
\end{eqnarray}
where $\rho_{ji} ({\bf{H}}^{(m)}, \widetilde {\bf{V}}_{ki}^A, {\bf{V}}_{ki}^E, \sigma_{w,i}^2) = \lambda_{ji}^{E \, (m)} {\bf{u}}_{ji}^{E \, (m)} {\bf{V}}_{ki}^E {\bf{u}}^{(m) \, \dagger}_{ji} ( {\bf{h}}_{ji}^{A \, (m)} \widetilde {\bf{V}}_{ki}^{A} {\bf{h}}_{ji}^{A  \, (m) \, \dagger} + \sigma_{w,i}^2 ||{\bf{h}}_{ji}^{A \, (m)}||^2   )$. At the $n$-th iteration, ${\bf{V}}^{E \, (n)}$ and $\{R_j^{(n)}\}$ are obtained by solving the following optimization problem based on SSUM method 
\begin{subequations} \label{OP_wSSUM_CBP}
\begin{eqnarray} 
\hspace{-0.5cm} {\bf{V}}^{E \, (n)}, \{R_j^{(n)}\} \gets && \hspace{-0.5cm} \underset {{\bf{V}}^{E}, \{R_j\}}{\textrm{arg max}} \sum_{j \in \mathcal{N}_M} \hspace{-0.1cm} R_j \label{OF_OP_wSSUM;CBP} \\
\textrm{s.t.} && \hspace{0.2cm}  R_j \le \frac{1}{n} \sum_{m=1}^{n} \widetilde R_j ({\bf{H}}^{(m)}, {\bf{V}}^E | {\bf{V}}^{E \, (m-1)}, \widetilde {\bf{V}}^{A \, (m)}, {\pmb{\sigma}}_{w}^{2 \, (m)}), \,\, \forall j \in \mathcal{N}_M, \label{RC_OP_wSSUM;CBP} \\
&& \hspace{0.2cm} C_{w,i} (\widetilde {\bf{V}}_i^A({\bf{H}}), {{\sigma}}_{w,i}^2({\bf{H}})) \le \bar C_i - \sum_{j \in \mathcal{M}_i} R_j, \label{FC_OP_wSSUM;CBP} \\
&& \hspace{0.2cm} P_i (\widetilde {\bf{V}}_i^A({\bf{H}}), {\bf{V}}_i^{E}, {{\sigma}}_{w,i}^2({\bf{H}})) \le \bar P_i, \hspace{1cm} \forall i \in \mathcal{N}_R \label{PC_OP_wSSUM;CBP}
\end{eqnarray}
\end{subequations}
until convergence.
\newpage
\bibliographystyle{IEEEtran}
\bibliography{refKJK}

\end{document}